\documentclass[preprint2,twocolumn,times,tighten]{aastex63}

\usepackage{color}
\usepackage{enumitem}
\usepackage{hyperref}
\usepackage{verbatim}
\usepackage{amsmath,amstext,mathrsfs}
\usepackage[all]{hypcap} 
\usepackage{afterpage}    
\usepackage{marginnote}
\usepackage{makecell}
\usepackage{subfigure}
\usepackage{multirow}

\shorttitle{Study Turbulence and Probe Magnetic Field Using Gradients Technique}
\shortauthors{Hu, Lazarian \& Bialy}
\graphicspath{{./}{figures/}}

\begin{document}
	
	\title{Study Turbulence and Probe Magnetic Fields Using Gradients Technique: Application to \ion{H}{1}-to-H$_2$ Transition Regions}

	\email{yue.hu@wisc.edu; alazarian@facstaff.wisc.edu; sbialy@cfa.harvard.edu}
	
	\author[0000-0002-8455-0805]{Yue Hu}
	\affiliation{Department of Physics, University of Wisconsin-Madison, Madison, WI 53706, USA}
	\affiliation{Department of Astronomy, University of Wisconsin-Madison, Madison, WI 53706, USA}
	\author{A. Lazarian}
	\affiliation{Department of Astronomy, University of Wisconsin-Madison, Madison, WI 53706, USA}
	\affil{Center for Computation Astrophysics, Flatiron Institute, 162 5th Ave, New York, NY 10010}
	\author[0000-0002-0404-003X]{Shmuel Bialy}
	\affiliation{Harvard-Smithsonian Center for Astrophysics, 60 Garden street, Cambridge,
		MA, USA}
	
	\begin{abstract}
		The atomic-to-molecular (\ion{H}{1}-to-H$_2$) transition in photodissociation regions (PDRs) has been investigated over the last several decades through analytic and numerical modeling. However, classical PDR models typically assume uniform density gas, ignoring the turbulent nature of the interstellar medium.
		Recently, \citet{2017ApJ...843...92B,2019ApJ...885..109B} have presented a theoretical framework for studying the \ion{H}{1}-to-H$_2$ in a realistic turbulent medium with a non-homogeneous density structure. Here we extend these turbulent-chemical models to explore the possibility of tracing the magnetic field direction in turbulent PDRs using the Gradients Technique.
		We utilize both subsonic and supersonic magnetohydrodynamic numerical simulations for chemical \ion{H}{1}/H$_2$ balance calculations. 
		We confirm that the density fluctuations induced by turbulence can disperse the distribution of H$_2$ and \ion{H}{1} fraction. We find that the energy spectrum of moment maps gets shallower when the sonic Mach number $\rm M_S$ increases. We explore the ability in magnetic field tracing of gradients of higher-order velocity centroids and compare their performance with that of traditional velocity centroid gradients (VCGs) and with intensity gradients (IGs). We find that the velocity gradients of the second-order centroids (VC$_2$Gs) are more accurate than VCGs and IGs in probing the magnetic field orientation.
	\end{abstract}
	
	
	\section{Introduction}
	\label{sec:intro}
	Photodissociation regions (PDRs) are regions in the interstellar space, typically at the boundaries of molecular clouds,  where the far ultraviolet (FUV) radiation plays a dominant role in photodissociating molecules and heating the gas \citep{1999RvMP...71..173H}. Owing to dust absorption and H$_2$ self-shielding the FUV is attenuated with increasing depth into a PDR and the gas undergoes a transition from atomic to molecular (\ion{H}{1}-to-H$_2$) form \citep[e.g., ][]{2007ApJ...654..273G, 2008ApJ...689..865K,2009ApJ...690.1497H,2014ApJ...790...10S,Bialy2015_perseus, Bialy2017_w43,2020MNRAS.492L..45B}. 
	
	The conversion to molecular form is critical for the formation of other molecules that are used as important observational tracers of the ISM \citep{1973ApJ...185..505H,1995ApJS...99..565S,2013RvMP...85.1021T,2013ChRv..113.9043V,2015MNRAS.450.4424B}. The PDRs in the interstellar medium (ISM) are threaded by both magnetic field and turbulence \citep{1981MNRAS.194..809L,2004ARA&A..42..211E, 2007prpl.conf...63B,MO07,2010ApJ...710..853C}. There, magnetic field and turbulence play crucial roles in regulating molecules' formation, controlling the heat transfer, and constraining star formation \citep{1965ApJ...142..584P,1966ApJ...146..480J, 1979cmft.book.....P, 2011Natur.479..499L,2013ApJ...768..159H, 2014ApJ...783...91C}. For instance, the high-density contrast and filaments created by supersonic turbulence serve as the nurseries for new stars \citep{2004RvMP...76..125M,MO07,2010A&A...512A..81F,Hu20}. 
	
	In addition to turbulence, observational measurement of the magnetic field presents a big challenge. The primary ways to probe the magnetic fields in ISM include the measurements of polarized dust emission \citep{Aetal15,2016ApJ...824..134F}, stellar light polarization \citep{2000AJ....119..923H,2015MNRAS.452..715P}, molecular-line splitting Zeeman effect \citep{2010ApJ...725..466C,2012ARAA...50..29}, and the Faraday rotation \citep{2012A&A...542A..93O}. There are, however, difficulties when studying the magnetic field through dust polarimetry. For one, the stellar light polarization can only sample the magnetic fields in the direction toward stars producing discrete magnetic field morphology \citep{1970MmRAS..74..139M,1990ApJ...359..363G,2019A&A...624L...8P}. Also, the measurement of dust polarimetry is generally difficult, since the grain alignment efficiency drops significantly in the case of high optical depth, limiting the reliability of tracing the magnetic field in optically thick regions \citep{ 2007ApJ...669L..77L, Aetal15}. The Zeeman measurement only gives the signed magnetic field strength along the line of sight (LOS) and requires exceptionally high sensitivity and long integration times. As for Faraday rotation, it measures the electron density-weighted magnetic field strength along the LOS and, therefore, generally does not probe the magnetic field in primarily neutral regions such as molecular clouds.
	
	Nevertheless, the Gradients Technique (GT) has been developed as a promising way for studying the magnetic fields across multiple scales in ISM  \citep{2017ApJ...835...41G,YL17a,LY18a, PCA}. GT utilizes the advancements of MHD turbulence theory and turbulent reconnection \citep{GS95,LV99}, in particular, the prediction that turbulent eddies are elongated in the direction of the magnetic field surrounding the eddy, i.e., the {\it local} magnetic field. 
	This theoretical prediction that is reliably supported by numerical simulations \citep{2001ApJ...554.1175M,2000ApJ...539..273C}. \cite{2002ApJ...564..291C} entails the conclusion that both gradients of density and velocity amplitude are perpendicular to the local direction of the magnetic field \footnote{The symmetry of Alfv\'{e}nic perturbations in terms of velocity and magnetic field entails that the magnetic fluctuations also share the same property. This gave rise to the development of the magnetic field tracing based on synchrotron intensity gradients (SIGs) \citep{2018ApJ...855...72L} and synchrotron polarization gradients (SPGs) \citep{2018ApJ...865...59L}. We do not discuss these promising techniques in this paper.}. 
	Therefore, the magnetic fields can be probed by rotating the gradients with 90$^ \circ $ using spectroscopic information. 
	
	GT has been widely tested in numerical simulation and observation, spanning from diffuse transparent atomic gas \citep{2019ApJ...874...25G,EB} to molecular self-absorbing dense gas \citep{2019ApJ...873...16H,survey,velac,2020arXiv200715344A,PDF}. Several studies also extend the GT to estimate the magnetization level \citep{2018ApJ...865...46L} and the sonic Mach number \citep{2018arXiv180200024Y}, distinguishing shocks \citep{IGs}, and identifying the self-gravitating regions in molecular clouds \citep{Hu20}. GT can employ either the density gradient or velocity gradient. However, the statistics of density in super-sonic turbulence can significantly differ from velocities \citep{2005ApJ...624L..93B,2007ApJ...658..423K}. Nevertheless, the study in \citet{2005ApJ...624L..93B} revealed that for most of the super-sonic turbulent volume, the density passively follows the velocity fluctuations, the notable exceptions being shocks. Thus, we expect that for subsonic as well as for a significant portion of the super-sonic turbulent volume, the density gradients to behave similarly to the gradients of velocity. This constitutes the theoretical justification of intensity gradients (IGs; \citealt{IGs}) that we also employ in this paper.\footnote{The IGs should not be confused by the Histograms of Relative Orientation (HRO) proposed by \citet{Soler2013}. While both techniques employ intensity gradients, the IGs use the set of procedures from the VGT to obtain the magnetic field direction. On the contrary, the HRO gets the magnetic field via polarization measurements and compares those with the magnetic field in order to determine a critical density at which the change of the gradients and polarization occurs in molecular clouds. A detailed comparison of IGs and HRO is presented in \citet{IGs}.}
	
	The challenge that we address here is related to the application of the GT to PDR. For instance, for the PDRs at the cloud boundaries, the anisotropic properties of turbulence might be distorted by the transition from atomic gas to molecular gas. Therefore, it is essential to study the effect of turbulence on the chemical structure of interstellar clouds and the performance of GT in PDRs. In this work, we focus on the turbulence's properties of \ion{H}{1}-to-H$_2$ transition produced by photodissociation at the cloud boundaries \citep{2017ApJ...843...92B,2019ApJ...885..109B}. 
	We generate synthetic \ion{H}{1} and H$_2$ cubes in chemical balance by post-processing numerical MHD simulations. 
	
	For extracting the density information, we employ the intensity map (i.e., moment-0 map, \citealt{YL17b,IGs}), the velocity centroid map (i.e., moment-1 map, \citealt{2017ApJ...835...41G, YL17a}) and the velocity channel map \citep{LY18a} to obtain the velocity information in spectroscopic data. In this work, we explore the ability of gradients of higher-order centroids to trace the magnetic field. In particular, we use the gradients of second-order velocity centroids, which we term VC$_2$Gs, to distinguish from the traditional velocity centroid gradients (VCGs). We use this modification to trace the magnetic field and make a comparison with the one obtained from moment-0 and moment-1 maps. 
	
	In what follows, we illustrate the theoretical foundation of the GT in terms of MHD turbulence in \S~\ref{sec:theory}. We give details about the analytical model of the \ion{H}{1}-to-H$_2$ transition and the numerical simulations used in this work in \S~\ref{sec:data}. In \S~\ref{sec:method}, we describe the full algorithm in probing the magnetic fields through GT. In \S~\ref{sec:result}, we analyze the properties of turbulence in PDRs through the energy spectrum and structure-function. We compare the ability of GT in tracing magnetic filed using different moment maps. In \S~\ref{sec:diss} and \S~\ref{sec:conc}, we give our discussion and conclusions.
	
	\section{Theoretical consideration}
	\label{sec:theory}
	
	\subsection{MHD turbulence theory}
	\label{subsec:MHD}
	\begin{figure*}[t]
		\centering
		\includegraphics[width=0.78\linewidth,height=0.5\linewidth]{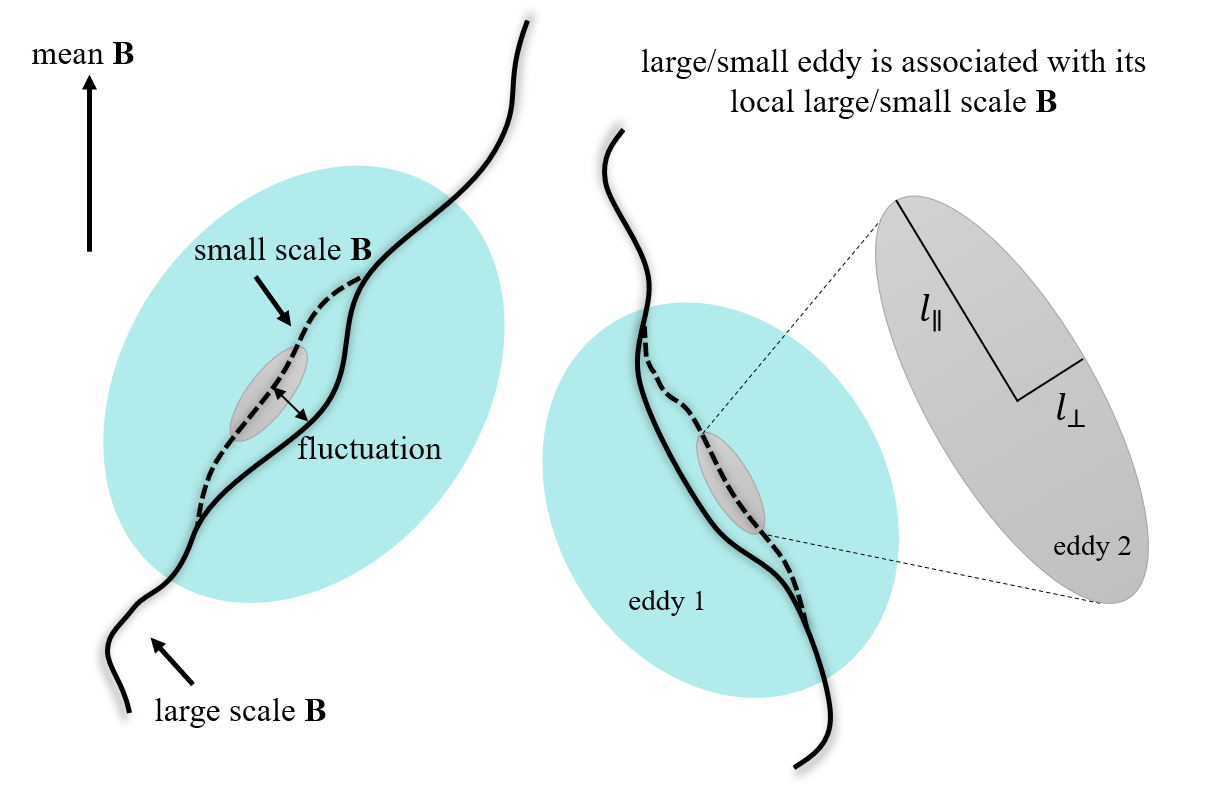}
		\caption{\label{fig:VGT} For trans-Alfv\'{e}nic turbulence, large eddy 1 is almost isotropic since they have similar semi-major  axes ($l_\parallel$) and  semi-minor axes ($l_\bot$). Smaller eddy 2 has a relatively larger semi-major axis to the semi-minor axis ratio. Therefore, they are relatively more elongated. The solid curve defines the directions of the local mean magnetic field line \textbf{B} for eddy 1, while the dashed curves define the directions of the local mean magnetic field line for eddy 2 and other small scale eddies. While large eddies induce the magnetic field's global change, the small eddies still follow the local magnetic field. Extracted from \citet{cluster}.}.
	\end{figure*}
	MHD turbulence theory is at the GT's foundations, and it capitalizes on the advancements of this theory. The understanding that MHD turbulence is anisotropic came through theoretical and numerical work several decades ago \citep{1981PhFl...24..825M,1983PhRvL..51.1484M,1983JPlPh..29..525S,1984ApJ...285..109H}. In this section, we briefly explain the basic elements of the theory that are essential for understanding the GT. An in-depth discussion of the properties of MHD turbulence can be found in the monograph by \citet{BL19}.
	\begin{figure*}[t]
		\centering
		\includegraphics[width=1.0\linewidth,height=0.6\linewidth]{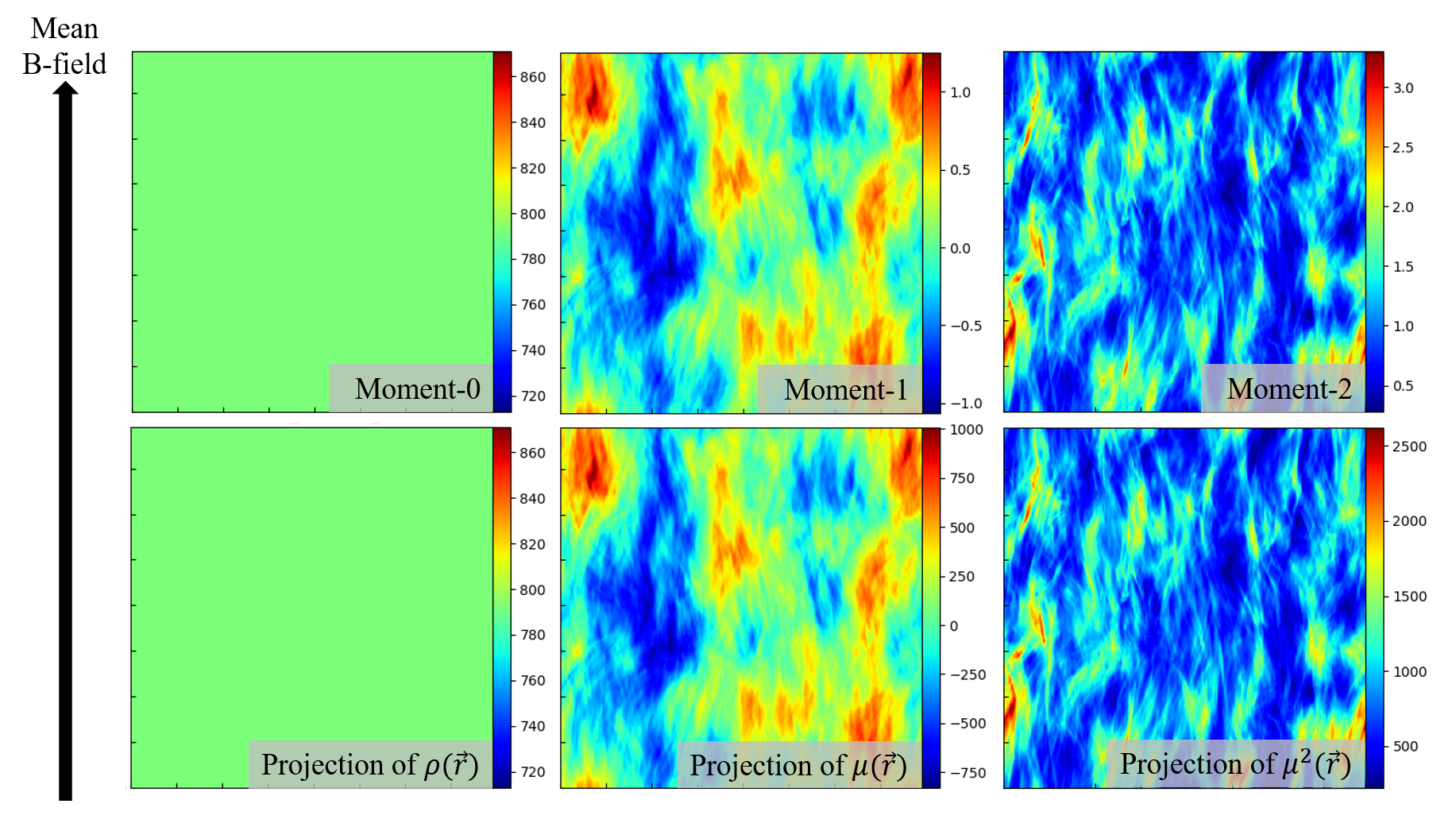}
		\caption{\label{fig:proj} \textbf{Top:} An example of three moment maps. The PPV cube is produced from the MHD simulation A3 using a uniform density field $\rho(\Vec{r})=1$ while keeping the original velocity field unchanged $\rm M_S$ = 10.54. \textbf{Bottom:} the projection of $\rho(\Vec{r})$, $\mu(\Vec{r})$, and $\mu^2(\Vec{r})$ along the LOS z-axis. $\mu(\Vec{r})$ is the LOS component of turbulent velocity.}
	\end{figure*}
	
	A crucial boost to the theory of MHD turbulence was given by \citet{GS95}, denoted as GS95 later. GS95 explicitly derived that the turbulence eddies are elongating along the magnetic field, showing the scaling relation:
	\begin{equation}
		k_\parallel\propto(k_\bot)^{2/3}
	\end{equation}
	where $k_\bot$ and $k_\parallel$ are wavenumbers perpendicular and parallel to the magnetic field, respectively. The cornerstone of the GS95 theory is the concept of a ”critical balance” between parallel and perpendicular timescales $k_\parallel V_A\sim k_\bot v_l$, here $v_l$ is the RMS speed of turbulence at the scale $l$ and V$_A$ is the Alfv\'{e}nic speed. However, the GS95 anisotropy scaling is derived in the global, i.e., mean magnetic field reference frame, in which the predicted scaling is not valid.  To understand the actual dynamics, it is useful to consider the eddy-based picture of strong MHD turbulence that is based on the concept of turbulent reconnection. \citet{LV99} (henceforth LV99)  demonstrated that the turbulent reconnection process is fast and the time scale for
	the reconnection of the corresponding eddies equal to the
	eddy turnover time. This result meant that turbulent motions were not constrained by fluid motion perpendicular to the local magnetic field. Therefore one can conclude that hydrodynamic-type eddy motions if not affected by magnetic field back-reaction if the eddy rotations are aligned with the direction of the magnetic field surrounding the eddy. 
	
	For our present, the latter is the fundamental property at the foundation of the gradient technique. In other words, the alignment of eddies with the local magnetic field means that if we detect eddy's preferential rotation direction, we can determine the magnetic field direction at the eddy location. Therefore using velocity gradients, one can map the magnetic field direction in turbulent media.     
	
	Continuing with our picture of MHD turbulence as a collection of eddies aligned with the magnetic field, we can state that turbulent cascade, which gets minimal resistance for the mixing perpendicular to the magnetic field, directs most of the cascading energy. In this case, the critical balance is the natural consequence of equality of the time of eddy turnover: $l_\bot/v_l$ and the period of the Alfv\'{e}n wave $l_\parallel/V_A$. Note, that we use notations $l_\|$ and $l_\bot$ and not wavenumber notation. This is to stress that the measurements are done in the local reference system and not the reference system related to the mean-field. 
	
	Incidentally, for the calculations in LV99 provided for Alfv\'{e}n Mach number $M_A<1$, where $M_A$ is the ratio of the injection velocity $V_{inj}$ to the Alfv\'{e}n speed $V_A$, the following relation between the parallel and perpendicular scales of the eddies:
	\begin{equation}
		\label{eq.lv99}
		l_\parallel\simeq L_{inj}(\frac{l_\bot}{L_{inj}})^{\frac{2}{3}}M_A^{-4/3}
	\end{equation}
	where $L_{inj}$ is the injection scale of turbulence, $l_\bot$ denotes the eddies' scale perpendicular to the magnetic field. Note that this scaling relation is measured in the eddies' local reference frame, rather than the mean magnetic field used in GS95. This universal scale-dependent anisotropy of Alfv\'{e}nic turbulence in the local magnetic field reference frame has been demonstrated in \citet{2000ApJ...539..273C,2002ApJ...564..291C,2001ApJ...554.1175M}. Indeed, the local system of reference is critical for understanding the GT. For tracing the magnetic field using either density or velocity field, it is essential that the density fluctuation or velocity fluctuation is oriented in respect to the local magnetic field's direction, instead of the global mean magnetic field, see Fig.~\ref{fig:VGT}. 
	
	Combining Eq.~\ref{eq.lv99} and the "critical balance" expressed in the local reference frame, i.e., $l_\bot V_A\sim l_\parallel v_l$, one can get the scaling relation for velocity fluctuations (see LV99):
	\begin{equation}
		\label{eq:KS}
		v_{l}\simeq v_L(\frac{l_{\perp}}{L_{inj}})^{\frac{1}{3}}M_A^{\frac{1}{3}}
	\end{equation}
	where $v_L$ is the injection velocity of turbulence. By taking square on both sides, the Kolmogorov's scaling relation $v_l^2\propto (l_\bot)^{2/3}$ is recovered. 
	
	The fluctuation induced by turbulence is more complicated in terms of the stochastic density field. \citet{2003MNRAS.345..325C} derived that the density $\rho_k$ of the eddy at scale $k$ in Fourier space can be expressed as:
	\begin{equation}
		|\rho_k|=\frac{\rho_0v_k}{c}|\hat{k}\cdot\hat{\zeta}|
	\end{equation}
	where $\rho_0$ is the mean density, $\hat{\zeta}$ is the unit vector for the Alfv\'{e}nic mode, fast mode, or slow mode, c is the  propagation speed of corresponding mode, and $v_k$ is the turbulence's velocity at scale $k$ in Fourier space. The density fluctuation $\rho_l$ in real space is therefore obtained from the inverse Fourier transform:
	\begin{equation}
		\rho_l=\mathscr{F}^{-1}( |\rho_k|)=\frac{\rho_0v_l}{c}\mathscr{F}^{-1}(|\hat{k}\cdot\hat{\zeta}|)
	\end{equation}
	Explicitly, since the anisotropic relation indicates $l_\bot \ll l_\parallel$, the velocity gradient and density gradient scale as \citep{2018arXiv180200024Y,cluster}:
	\begin{equation}
		\label{eq.grad}
		\begin{aligned}       
			\nabla \rho_l&\propto\frac{\rho_{l}}{l_\bot}\simeq\frac{\rho_0}{c}\mathscr{F}^{-1}(|\hat{k}\cdot\hat{\zeta}|)\nabla v_l\\
			\nabla v_l&\propto\frac{v_{l}}{l_\bot}\simeq \frac{v_L}{L_{inj}}(\frac{l_{\perp}}{L_{inj}})^{-\frac{2}{3}}M_A^{\frac{1}{3}}\\
			\nabla v_l^2&\propto\frac{v_{l}^2}{l_\bot}\simeq \frac{v_L^2}{L_{inj}}(\frac{l_{\perp}}{L_{inj}})^{-\frac{1}{3}}M_A^{\frac{2}{3}}\\
		\end{aligned}
	\end{equation}
	The gradients induced by both velocity and density fluctuations increase respectively as $v_l/l_\bot\sim l^{-2/3}$ and $\rho_l/l_\bot\sim l^{-2/3}$. The direction of gradients is perpendicular to the local directions of magnetic field
	\citep{2001ApJ...554.1175M,2000ApJ...539..273C}. Also, this means that the smallest eddies resolved in observations provide the most important contribution for the gradients. This consideration is at the core of GT.
	
	\subsection{Gradient of moment maps}
	In observation, the gas distribution  in a
	given spectral line is defined in Position-Position-Velocity (PPV) cubes toward some direction on the sky and at a given LOS velocity $v$, rather than in the real-space $\Vec{r}=(x, y, z)$. The LOS component of velocity $v$ at the position $(x,y)$ is a sum of the turbulent velocity $\mu(\Vec{r})$ and the residual component due to thermal motions. This residual thermal velocity $v-\mu(\Vec{r})$ has a Maxwellian distribution:
	\begin{equation}
		\label{eq.max}
		\Phi(v-\mu(\Vec{r}))=\frac{1}{\sqrt{2\pi\beta(\Vec{r})}}\exp[-\frac{(v-\mu(\Vec{r}))^2}{2\beta(\Vec{r})^2}]
	\end{equation}
	where $\beta(\Vec{r})=k_BT(\Vec{r})/m$, $m$ being the mass of atoms. The temperature $T(\Vec{r})$ can vary from point to point if the emitter is not isothermal.
	
	The relation between 
	the gas distribution $\rho(x,y,v)$ in PPV cubes and $\rho(\Vec{r})$ in real-space is given by \citep{2004ApJ...616..943L}:
	\begin{equation}
		\rho(x,y,v)=\int\rho(\Vec{r})\Phi(v-\mu(\Vec{r}))dz
	\end{equation}
	where $\Phi(v-\mu(\Vec{r}))$ is the Maxwell’s distribution of the thermal component of LOS velocity, and $\mu(\Vec{r})$ is the LOS turbulent velocity.
	
	For the gas distribution $\rho(x,y,v)$, we can define the moment-0 map $f_0(x,y)$ (i.e., the intensity map) and the moment-n map $f_n(x,y)$ as:
	
	\begin{equation}
		\begin{aligned}
			f_0(x,y)&=\int\rho(x,y,v)dv=\int\rho(\Vec{r})dz\int\Phi(v-\mu(\Vec{r}))dv\\&=\int\rho(\Vec{r})dz\\
			f_n(x,y)&=\frac{\int v^n\rho(x,y,v)dv}{\int\rho(x,y,v)dv},n\ge1\\
		\end{aligned}
	\end{equation}
	
	The moment corresponding to $n=1$ is called " velocity centroid," representing the averaged line-of-sight velocity when density is constant. The constructions with $n>1$ we will call "higher moment centroids."
	
	Applying the gradient operator $\nabla_{2D}=(\frac{\partial}{\partial x},\frac{\partial}{\partial y},0)^T$ to $f_0(x,y)$, the gradient amplitude can be expressed as \citep{2018arXiv180200024Y}:
	\begin{equation}
		\begin{aligned}
			|\nabla_{2D} f_0(x,y)|&\propto|\int \nabla_{2D}\rho(\Vec{r}) dz|\\
			&=\langle|\nabla_{3D}\rho(\Vec{r})|\cos\gamma\rangle\sqrt{N L_{inj}^2}
		\end{aligned}
	\end{equation}
	in which we assume there are $N = L/L_{inj}$ eddies along LOS with distance $L$ and the density gradient follows a random walk summation. $\langle...\rangle$ denotes the average value, 3D gradient operator is $\nabla_{3D}=(\frac{\partial}{\partial x},\frac{\partial}{\partial y},\frac{\partial}{\partial z})^T$, and $\gamma$ is the relative angle between the gradient and the POS with $\cos\gamma=|\nabla_{2D}\rho(\Vec{r})|/|\nabla_{3D}\rho(\Vec{r})|$. Note this summation of gradient amplitude as random walk is only valid when tan$(\gamma/4)>M_A/\sqrt{3}$ \citep{2020arXiv200207996L}. The direction of the gradient is then:
	\begin{equation}
		\label{eq.gradf0}
		\begin{aligned}
			\nabla_{2D} f_0(x,y)&=\textbf{U}\cdot\sum_{i=1}^N\nabla_{3D}\rho_i(\Vec{r})
		\end{aligned}
	\end{equation}
	where \textbf{U} is the projection operator:
	$$
	\textbf{U}=\begin{pmatrix} 
		+1 & 0 & 0 \\
		0 & +1 & 0 \\
		0 & 0 & 0
	\end{pmatrix}
	$$
	The direction of $\nabla_{2D} f_0(x,y)$ is therefore the vector summary of the density gradient along LOS. As for the moment-n map $f_n(x,y)$, in the limit of constant density distribution, $f_n(x,y)$ provides the value of $\mu(\Vec{r})^n$ averaged along the line-of-sight. In particular, for moment-1 and moment-2 maps, we have:
	\begin{equation}
		\label{eq.12}
		\begin{aligned}
			f_1(x,y)&=\frac{\int\rho(\Vec{r})dz\int v\Phi(v-\mu(\Vec{r}))dv}{\int \rho(z)dz}\\
			&\propto\frac{\int\rho(\Vec{r})\mu(\Vec{r})dz}{\int \rho(\Vec{r})dz}\\
			f_2(x,y)&=\frac{\int\rho(\Vec{r})dz\int v^2\Phi(v-\mu(\Vec{r}))dv}{\int \rho(\Vec{r})dz}\\
			&\propto\frac{\int\rho(\Vec{r})\mu^2(\Vec{r})dz}{\int \rho(\Vec{r})dz}\\
		\end{aligned}
	\end{equation}
	Note if we consider also the regular gas flow due to Galactic rotation $v_{gal}(\Vec{r})$. The residual thermal velocity becomes $v-v_{gal}(\Vec{r})-\mu(\Vec{r})$ which has a Maxwellian distribution.
	This introduces one more velocity term $\frac{\int\rho(\Vec{r})v_{gal}(\Vec{r})dz}{\int \rho(\Vec{r})dz}$ to $f_1(x,y)$ and $\frac{\int\rho(\Vec{r})v_{gal}^2(\Vec{r})dz}{\int \rho(\Vec{r})dz}$ to $f_2(x,y)$, which can be considered as constant for an individual cloud due to the little variation in $v_{gal}(\Vec{r})$. 
	
	As shown in Fig.~\ref{fig:proj}, we produce a synthetic PPV cube from the MHD simulation A3 (see \S~\ref{sec:data} for details), using a unity density field while keeping the original velocity field  $\rm M_S$ = 10.54 unchanged. We calculate three moment maps from the PPV cube and project the density field $\rho(\Vec{r})$ and velocity field $\mu(\Vec{r})$ along the z-axis. We find that moment maps exhibit very similar structures with the projected density and velocity maps. In particular, these structures are elongating along the magnetic field. This agrees with our theoretical consideration. In the rest of the paper, when we study the various moment projections in \ion{H}{1}-to-H$_2$ gas, we take into account both fluctuations in the density as well as in the velocity.
	
	The gradients of corresponding $f_1(x,y)$ and $f_2(x,y)$ can be written as:
	\begin{equation}
		\label{eq.f12}
		\begin{aligned}
			\nabla_{2D} f_1(x,y)&\propto\frac{\int\nabla_{2D}[\rho(\Vec{r})\mu(\Vec{r})]dz-f_1(x,y)\int\nabla_{2D}\rho(\Vec{r})dz}{f_0}\\
			&=\frac{\textbf{U}}{f_0}\cdot\sum_{i=1}^N[(\mu_i-f_1)\nabla_{3D}\rho_i+\rho_i\nabla_{3D}\mu_i]\\
			\nabla_{2D} f_2(x,y)&\propto\frac{\int\nabla_{2D}[\rho(\Vec{r})\mu^2(\Vec{r})]dz-f_2(x,y)\int\nabla_{2D}\rho(\Vec{r})dz}{f_0}\\
			&=\frac{\textbf{U}}{f_0}\cdot\sum_{i=1}^N[(\mu_i^2-f_2)\nabla_{3D}\rho_i+\rho_i\nabla_{3D}\mu_i^2]\\
		\end{aligned}
	\end{equation}
	The subscript $i$ indicates the variable of the i-th eddy. The direction of $\nabla_{2D} f_1(x,y)$ and $\nabla_{2D} f_2(x,y)$ consist of the contribution from both density and velocity field. In the limit of constant density distribution, the density gradient can be erased but only the velocity gradient gives contribution. Note in the case of large scale study in which $v_{gal}(\Vec{r})$ cannot be assumed as constant, one has to consider the gradients of the $v_{gal}(\Vec{r})$ term. Nevertheless, considering the flux freezing condition, the magnetic field is expected to follow the radial direction of regular gas flow caused by Galactic rotation. Coincidentally, the gradients, in this case, are still perpendicular to the magnetic field showing a different scaling relation from Eq.~\ref{eq.grad}. Similarly, the gradient amplitude is summed in a random walk manner:
	\begin{equation}
		\begin{aligned}
			|\nabla_{2D} f_1(x,y)|&\propto\frac{\sqrt{NL_{inj}^2}}{f_0}\langle|(\mu-f_1)\nabla_{3D}\rho+\rho\nabla_{3D}\mu|\cos\gamma\rangle\\
			|\nabla_{2D} f_2(x,y)|&\propto\frac{\sqrt{NL_{inj}^2}}{f_0}\langle|(\mu^2-f_2)\nabla_{3D}\rho+\rho\nabla_{3D}\mu^2|\cos\gamma\rangle\\
		\end{aligned}
	\end{equation}
	Note here we assume the density gradient and velocity gradient exhibits the same relative angle $\gamma$ with respect to POS, as both gradients are perpendicular to the magnetic field. Following similar steps, we can extend the calculation to the un-normalized moment maps $F_n(x,y)$:
	\begin{equation}
		\label{eq.14}
		\begin{aligned}
			&F_1(x,y)=\int v\rho(x,y,v)dv\propto\int\rho(\Vec{r})\mu(\Vec{r})dz\\
			&F_2(x,y)=\int v^2\rho(x,y,v)dv\propto\int\rho(\Vec{r})\mu^2(\Vec{r})dz\\
			&\nabla_{2D}F_1(x,y)\propto\textbf{U}\cdot\sum_{i=1}^N(\mu_i\nabla_{3D}\rho_i+\rho_i\nabla_{3D}\mu_i)\\
			&\nabla_{2D}F_2(x,y)\propto\textbf{U}\cdot\sum_{i=1}^N(\mu_i^2\nabla_{3D}\rho_i+\rho_i\nabla_{3D}\mu_i^2)\\
			&|\nabla_{2D}F_1(x,y)|\propto\sqrt{NL_{inj}^2}\langle|\mu\nabla_{3D}\rho+\rho\nabla_{3D}\mu|cos\gamma\rangle\\
			&|\nabla_{2D}F_2(x,y)|\propto\sqrt{NL_{inj}^2}\langle|\mu^2\nabla_{3D}\rho+\rho\nabla_{3D}\mu^2|cos\gamma\rangle\\
		\end{aligned}
	\end{equation}
	\begin{table}
		\centering
		\label{tab:sim}
		\begin{tabular}{| c | c | c | c | c |}
			\hline
			Model & $\rm M_S$ & M$_A$  & Resolution & $\beta$\\ \hline \hline
			A0 & 0.66 & 0.12 & $792^3$ & 0.066\\  
			A1 & 1.27 & 0.50 & $792^3$ & 0.310\\
			A2 & 5.64 & 0.31 & $792^3$ & 0.006\\  
			A3 & 10.54 & 0.46 & $792^3$ & 0.004\\  
			\hline
		\end{tabular}
		\caption{Description of our MHD simulations. $\rm M_S$ and M$_A$ are the instantaneous values at each the snapshots are taken. The compressibility of turbulence is characterized by $\beta=2(\frac{\rm M_A}{\rm M_S})^2$.}
	\end{table}
	
	Comparing with Eq.~\ref{eq.f12}, we can see the magnitude of density gradient is suppressed in the normalized moment maps. Nevertheless, as we discussed in \S~\ref{subsec:MHD}, for incompressible turbulence, the scaling of density gradient is similar to that of the velocity gradient (see Eq.~\ref{eq.grad}). Therefore, we can expect that the gradient of moment maps exhibits exactly the velocity gradient properties in the incompressible limit, i.e., the direction of moment maps' gradients is expected to be perpendicular to the projected magnetic field.
	
	A particular case for GT is the presence of strong shocks. The density gradient is preferentially perpendicular to the shock front due to the rapid jump of density and velocity. Coincidentally, the magnetic field tends to be perpendicular to the shock front as well in ISM \citep{2019ApJ...878..157X}. It results from the shocks wave propagating along the magnetic field. The magnetic field suppresses the compression in the perpendicular direction so that the fluid flows the magnetic field line. Therefore, the shock compression in supersonic turbulence produces the perpendicular structures. Consequently, the density gradient becomes parallel to the magnetic field, instead of being perpendicular \citep{YL17b,IGs}. As for the velocity gradient, the situation gets different. As shown in Fig.~\ref{fig:proj}, the projected supersonic velocity field is still parallel to the magnetic field, rather than being perpendicular. The corresponding velocity gradient is always perpendicular to the magnetic field in both the subsonic and supersonic environment. 
	
	We know that $\nabla_{2D}f_0(x,y)$ consists of only density gradient, see Eq.~\ref{eq.gradf0}. The gradient of moment-0 is, therefore, sensitive to shocks. However, from Eq.~\ref{eq.f12}, we know that $\nabla_{2D}f_1(x,y)$ and $\nabla_{2D}f_2(x,y)$ are projected from the summation of 3D velocity and density gradients along LOS. To see this change of gradient's direction requires the shock to be strong enough so that the density's contribution gets dominated.
	
	\section{Numerical Data}
	\label{sec:data}
	
	\subsection{MHD simulations}
	We perform 3D MHD simulations through ZEUS-MP/3D code \citep{2006ApJS..165..188H}, which solves the ideal MHD equations in a periodic box. We use single fluid, operator-split, solenoidal turbulence injections, and staggered grid MHD Eulerian assumption. To emulate a part of an interstellar cloud, we use the barotropic equation of state, i.e., these clouds are isothermal with temperature T = 50.0 K, sound speed $c_s$ = 0.42 km/s and cloud size L = 10 pc. The sound crossing time $t_v = L/c_s$ is $\sim 23.2$ Myr, which is fixed owing to the isothermal equation of state. This allows a natural extension of the isothermal \ion{H}{1}-to-H$_2$ transition model proposed in \citet{2014ApJ...790...10S} and \citet{Bialy2016_tran}, into the turbulent regime, similarly to the \citet[][]{2017ApJ...843...92B, 2019ApJ...885..109B} turbu-chemical model.
	
	MHD turbulence is characterized by sonic Mach number M$_{s}=v_{l}/c_{s}$ and Alfv\'{e}nic Mach numbers M$_{A}=v_{l}/v_{A}$, where $v_{l}$ is the injection velocity and $v_{A}$ is the Alfv\'{e}nic velocity. The turbulence is highly magnetized when the plasma's magnetic pressure is larger than the thermal pressure, i.e., M$_{A}<1$. The compressibility of turbulence is characterized by $\beta=2(\frac{\rm M_A}{\rm M_S})^2$. We refer to the simulations in Tab.~\ref{tab:sim} by their model name. For instance, our figures or captions will have the model name indicating which data cube is used to plot the figure. As indicated in Tab.~\ref{tab:sim}, all the simulations are highly compressible, so that the supersonic simulations ($\rm M_s$>1) develop strong density fluctuations. As we discuss below,  these density fluctuations result in fluctuations in the abundances of \ion{H}{1} and H $ _2$ in the turbulent boxes.

	\subsection{\ion{H}{1} and H$_2$ chemical balance}
	
	At any cloud depth and for unidirectional radiation normal to the cloud surface, the H$_2$ formation-destruction stead-state balance is given by:
	\begin{equation}
		\label{eq: H-H2 balance}
		Rn x_{\rm HI}=(\frac{D_0}{2}f_{shield}(N_{\rm H_2})e^{-\sigma_g N}+\zeta) x_{\rm H_2}
	\end{equation}
	The left hand side represent H$_2$ formation out of atomic H, where
	$R$ (cm$^3$ s$^{-1}$) is the H$_2$ formation rate coefficient,
	$n= n_{\rm H}+2n_{\rm H_2}$ is the hydrogen nuclei volume density, and $x_{\rm HI} \equiv n_{\rm HI}/n$ and $x_{\rm H_2} \equiv n_{\rm H_2}/n$ are the H and H$_2$ relative abundances. 
	The right-hand side represents H$_2$ destruction.
	The first term accounts for destruction by photodissociation where $D_0$ ($s^{-1}$) is the free-space H$_2$ photodissociation rate, $f_{shield}$ is the H$_2$ self-shielding function that depends on the accumulated H$_2$ column density $\rm N_{H2}$ from cloud edge to the point of interest, and $e^{-\sigma N}$ is the dust absorption attenuation term, where $\sigma_g$ (cm$^2$) is the dust absorption cross section per hydrogen nucleus integrated over the Lyman–Werner dissociation band (11.2–13.6 eV) and $\rm N=N_{HI}+2N_{H_2}$ is the total (atomic plus molecular) column density. 
	The factor of 1/2 accounts for the absorption of half the radiation by the optically thick slab. 
	The second term in the right-hand side of Eq.~(\ref{eq: H-H2 balance}) accounts for H$_2$ destruction by cosmic-rays (through ionization that is followed up by a series of abstraction reactions and by recombination that leads to the formation of H; \citealt{2015MNRAS.450.4424B}),  where $\zeta$ is the H$_2$ cosmic-ray ionization rate.
	Eq.~(\ref{eq: H-H2 balance}) is augmented by the mass conservation equation, $x_{\rm HI}+2 x_{\rm H_2}=1$.

	We calculate the \ion{H}{1} and H$_2$ abundances in all the simulations boxes in post-process, as follows. We use the density field from each simulation and calculate $x_{\rm HI}$ and $x_{\rm H_2}$ via Eq.~(\ref{eq: H-H2 balance}) assuming unidirectional UV field, and the standard parameter values: $D_0=5.8 \times 10^{-11}$ s${-1}$, $\zeta=2 \times 10^{-16}$ s$^{-1}$, $R=3 \times 10^{-17}$ cm$^{3}$ s$^{-1}$, and $\sigma_g=1.9\times 10^{-21}$ cm$^{-2}$. 
	These values correspond to the local mean UV radiation field \citep{1978ApJS...36..595D}, a standard dust to gas ratio, and H$_2$ formation rate in the cold neutral medium \citep{2014ApJ...790...10S}.
	For $f(N_{H_2})$ we use Eq.~(37) from \citet{1996ApJ...468..269D}.
	The H$_2$ abundances at different cloud depths are coupled through the H$_2$ self-shielding process.
	
	To derive the H and H$_2$ abundances in each location, we follow a similar procedure as in \citet[][]{2017ApJ...843...92B, 2019ApJ...885..109B}.  For each line-of-sight (i.e., a pair of coordinates, $(x,y)$), we solve Eq.~(\ref{eq: H-H2 balance}) starting from cloud edge ($z=0$), and progressively move inwards. At cloud edge, there is no attenuation, $f_{shield}(N_{H_2})e^{-\sigma_g N}=1$, and the \ion{H}{1} and H$_2$ abundances are readily given by Eq.~(\ref{eq: H-H2 balance}) (and the $x_{\rm HI}+2 x_{\rm H_2}=1$ constraint). We then make a (small logarithmic) step, $\Delta z$, into the cloud and calculate $x_{\rm HI}$ and $x_{\rm H_2}$ where now the shielding is calculated with $N_2 = n \Delta z x_{\rm H_2}$, $N=n \Delta z$. We continue this process until we reach the end of the simulation box, and repeat it for all the lines-of-sights. In each step, we interpolate the density as obtained from the MHD simulation onto the grid that is used for the calculation of the \ion{H}{1}-H$_2$ balance (which is much finer and is logarithmic). Finally, we interpolate back the resulting \ion{H}{1}-H$_2$ abundances back onto the native simulation grid. In the conclusion of this process, we have $x_{\rm HI}$ and $x_{\rm H_2}$ in each cell of each of the simulation boxes.
	
	With increasing distance from cloud edge, the UV radiation is attenuated via dust absorption and absorption in the H$_2$ lines (i.e., H$_2$ self-shielding), and the gas undergoes a transition from atomic to molecular state. Following \citet[][]{Bialy2016_tran}, for a uniform density gas, the \ion{H}{1}-to-H$_2$ transition occurs at a column density
	\begin{equation}
		\label{eq: Ntran}
		\rm N_{\rm tran} = 3.7\times10^{20} \ \ln \left[1+\left(\frac{n}{59 \  {\rm cm^{-3}}}\right )^{-0.7} \right] \ {\rm cm^{-2} } \ ,
	\end{equation}
	where we evaluated  Eqs.~22 and 39 of \citet[][]{Bialy2016_tran} assuming the standard parameter values. Because the \ion{H}{1}-H$_2$ balance depends on the gas density, in a supersonic turbulent medium, the transition  point varies between sightlines, due to the strong density fluctuations \citep[e.g., see Fig.~5 in][]{2017ApJ...843...92B}. Thus, Eq.~(\ref{eq: Ntran}) does not hold for supersonic turbulence, however, it is still useful as an analytic estimate for the typical depth into the simulation where the \ion{H}{1}-to-H$_2$ transition occurs (with $\rm n$ replaced with the mean density in the simulation, $\langle \rm n \rangle$). 
	
	In this work, we set $\langle \rm n \rangle=50 cm^{-3}$ for all simulations. The mean column density $\langle \rm N \rangle$ and $\rm N_{\rm tran}$ are therefore $\approx 1.40 \times10^{21} \rm cm^{-2}$ and $\approx 2.80 \times10^{20} \rm cm^{-2}$, respectively. Thus, on the one hand, the simulations have a sufficiently large column density to allow an \ion{H}{1}-to-H$_2$ transition for most of the sightlines, thus allowing us to study the statistics of both phases. On the other hand, the typical width of the transition zone is sufficiently large (with the simulation resolution of 780$^3$, the transition zone is of order $\sim 100$ cells), allowing us to resolve substructure in the \ion{H}{1} layers.
	
	\section{Methodology}
	\label{sec:method}
	\begin{figure}[t]
		\centering
		\includegraphics[width=1.0\linewidth,height=0.87\linewidth]{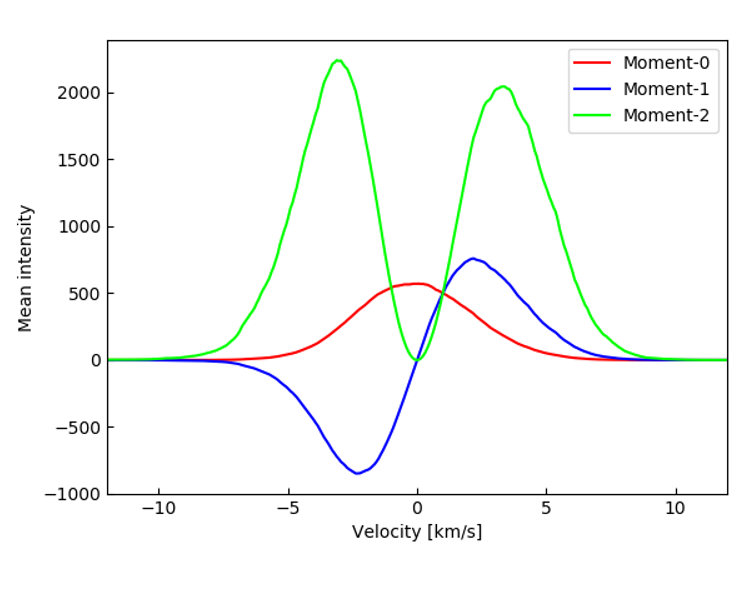}
		\caption{\label{fig:los} The spectral lines of moment-0 map (red), moment-1 map (blue), and  moment-2 map (green) as functions of LOS velocity. The spectral line is produced from simulation A3.}
	\end{figure}

	\subsection{Implementation of the Gradients Technique}
	The theory of intensity fluctuations in PPV space was pioneered in \citet{LP00} and later \citet{2008ApJ...686..350L} explored the possibility of using the statistics of intensity fluctuations in PPV cubes to study velocity turbulence. To detect the anisotropy of intensity and velocity fluctuations that are induced by the magnetic field in PPV cubes, GT usually employs a statistical description of either moment-0 map or moment-1 map \citep{2017ApJ...835...41G,YL17a,IGs}. 
	
	The moment-0 map $f_0(x,y)$ presents intensity map \textbf{I(x,y)} as it contains the information of intensity fluctuations. The normalized moment-1  map $f_1(x,y)$ reflects the velocity fluctuations and is denoted as, what we mentioned earlier, the velocity centroid map \textbf{C(x,y)}. The intensity map and normalized velocity centroid map are calculated as follow:
	\begin{equation}
		\begin{aligned}
			\rm I(x,y)&=f_0(x,y)=\int \rho(x,y,v)dv\\
			\rm C(x,y)&=f_1(x,y)=\frac{\int v\rho(x,y,v)dv}{\int \rho(x,y,v)dv}\\
		\end{aligned}
	\end{equation}
	where $\rho$ is the gas density, $v$ is the radial velocity along LOS. 
	The applicability of \textbf{I(x,y)} and \textbf{C(x,y)} in tracing magnetic field has been widely tested in both numerical simulation \citep{2017ApJ...835...41G,YL17a,IGs} and observation \citep{velac,cluster}. In this work, we further explore how to trace the magnetic field using the normalized moment-2 map, or second order centroids \textbf{S(x,y)}:
	\begin{equation}
		\begin{aligned}
			\rm S(x,y)&=f_2(x,y)=\frac{\int v^2\rho(x,y,v)dv}{\int \rho(x,y,v)dv}\\
		\end{aligned}
	\end{equation}
	
	In Fig.~\ref{fig:los}, we present the spectral line of each map along LOS using simulation A3. For moment-0, the spectral line as a function of the LOS velocity is simply the mean intensity in each corresponding channel. The spectral lines of 
	moment-1 and moment-2 are the mean intensity weighted by the corresponding LOS velocity and squared velocity, respectively. Given the spectral line, \textbf{I(x,y)} is effectively the integration over the entire line width carried out to find intensity. \textbf{C(x,y)} constructs the velocity-weighted moment of intensity over entire line width. Part of intensity information is removed, as the weighting can give the opposite sign. As for \textbf{S(x,y)}, the intensity of the central line becomes zeros through the weighting of squared velocity. It emphasis only the intensity in the surrounding of central lines.
	
	The pixelized gradient map $\psi_{g}(x,y)$, which denotes the angle of the gradients on the POS, of each moment map is calculated from the convolution of individual 2D map with 3 $\times$ 3 Sobel kernels $G_x$ and $G_y$\footnote{The Sobel kernels are defined as:
		$$
		G_x=\begin{pmatrix} 
			-1 & 0 & +1 \\
			-2 & 0 & +2 \\
			-1 & 0 & +1
		\end{pmatrix},
		G_y=\begin{pmatrix} 
			-1 & -2 & -1 \\
			0 & 0 & 0 \\
			+1 & +2 & +1
		\end{pmatrix}
		$$}:
	\begin{equation}
		\label{eq:conv}
		\begin{aligned}
			\bigtriangledown_x f(x,y)=G_x * f(x,y)  \\  
			\bigtriangledown_y f(x,y)=G_y * f(x,y)  \\
			\psi_{g}(x,y)=\tan^{-1}[\frac{\bigtriangledown_y f(x,y)}{\bigtriangledown_x f(x,y)}]
		\end{aligned}
	\end{equation}
	where $f(x,y)$ represents I(x,y), C(x,y), or S(x,y). $\bigtriangledown_x f(x,y)$ and $\bigtriangledown_y f(x,y)$ are the x and y components of gradient respectively. $*$ denotes the convolution. 
	
	Note that the perpendicular relative orientation of gradients and magnetic field only appears when the gradient sampling is enough, as the anisotropy of turbulent eddies concerning the local magnetic field is a statistical concept. Therefore, the raw gradient map $\psi_{g}(x,y)$ is not necessarily required to have any relation to the local magnetic field direction. The critical step after getting $\psi_{g}(x,y)$ is taking the average of gradients' orientation within a sub-block of interest, i.e., the sub-block averaging method, which is proposed by \citet{YL17a}. Because the gradients' orientation within a sub-block forms a Gaussian distribution,  the expectation value of the Gaussian distribution reflects the statistical most probable anisotropic gradient. Therefore, to probe the local magnetic field, we first draw the histogram of gradients' orientation and take the expectation value of the Gaussian fitting. By rotating the resulting gradient $\psi(x,y)$ with 90$^\circ$, we output the predicted the local mean magnetic field. Without specification, we automatically rotate the gradient by 90$^\circ$ to align with the magnetic field in the following.
	
	\subsection{Velocity fluctuation in thin velocity channel}
	In \S~\ref{sec:theory}, we derived that the moment-0 map consists of pure density contribution, while moment-1 and moment-2 maps can be used to obtain both density and velocity information. Nevertheless, \citet{LP00} presented an alternative way to extract the velocity fluctuations in PPV cubes through the effect of velocity caustics. There it was shown the velocity fluctuations are most prominent when the channel is sufficiently thin, i.e., the channel width $\Delta v$ satisfies:
	\begin{equation}
		\label{eq1}
		\begin{aligned}
			\Delta v<\sqrt{\delta v^2}, \thickspace\mbox{thin channel}\\
			\Delta v\ge\sqrt{\delta v^2},\thickspace\mbox{thick channel}\\
		\end{aligned}
	\end{equation} 
	where $\delta v$ is the velocity dispersion calculated from velocity centroid \citep{LP00}. Based on this theory, the thin velocity channel map has been employed to obtain velocity gradient for magnetic field tracing, which is named as Velocity Channel Gradients (VChGs, \citet{LY18a}). For VChGs, only the thin central channel \textbf{Ch(x,y)} is selected for calculation:
	\begin{equation}
		\begin{aligned}
			\rm Ch(x,y)=\int_{v_0-\Delta v/2}^{v_0+\Delta v/2}\rho(x,y,v) dv
		\end{aligned}
		\label{eq2}
	\end{equation}
	where $v_0$ is the velocity corresponding to the peak position in PPV's velocity profile. The validity of VChGs has been widely tested in both numerical and observational studies \citep{LY18a,survey,velac}. \citet{IGs} also gives a detailed comparison of IGs and VChGs. Later, instead of using only the thin central channel, \citet{EB} extend the calculation of velocity gradient to all thin channels in the PPV cube. This calculation introduces the pseudo-Stokes parameters and the Principal Component Analysis (PCA). The detailed recipe is presented in \citet{PCA} and \citet{EB}. Here we give only a brief review.  
	\begin{figure*}[t]
		\centering
		\includegraphics[width=1.00\linewidth,height=0.48\linewidth]{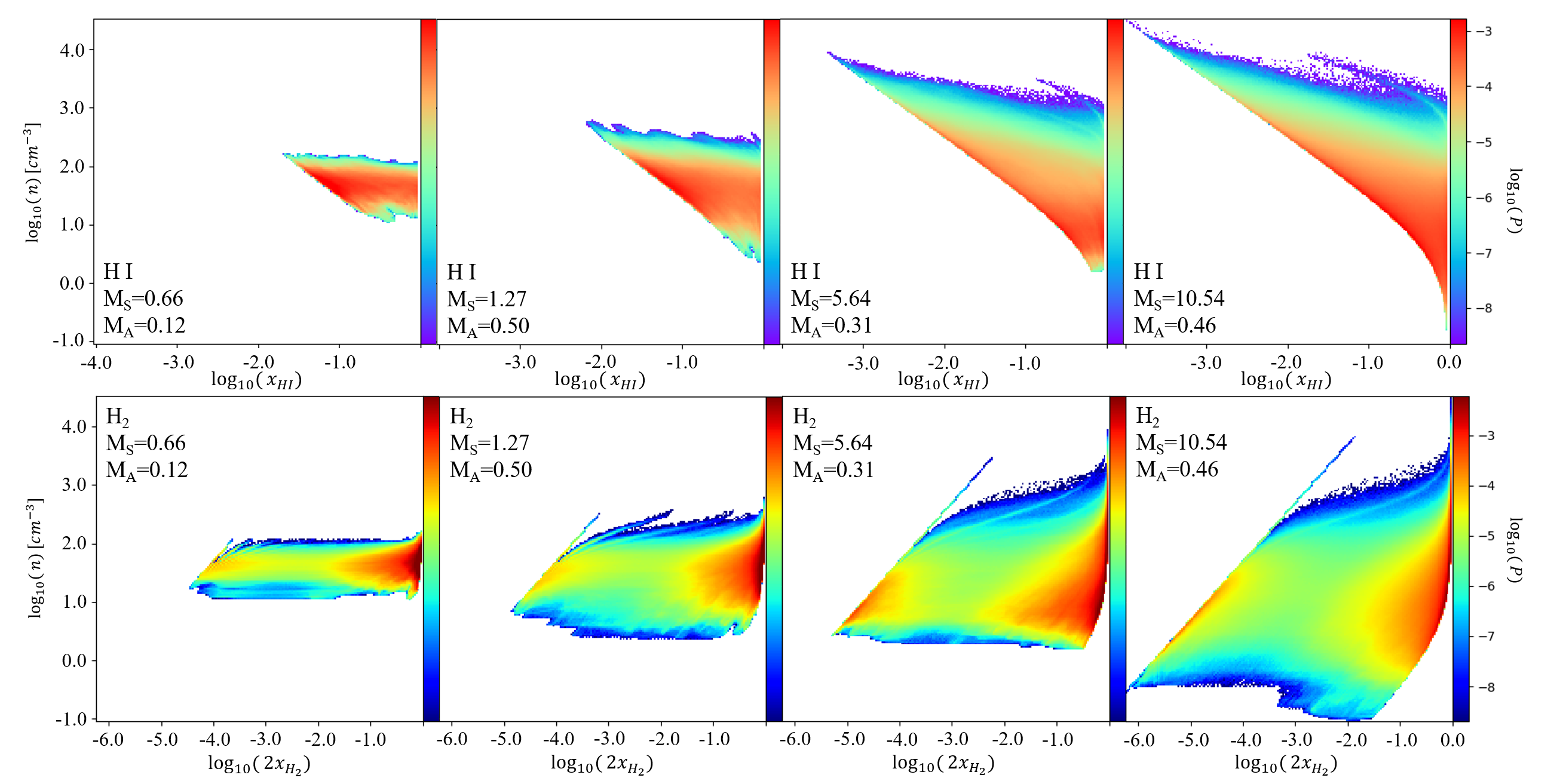}
		\caption{\label{fig:hist} 
			The 2D histograms of $\rm x_{HI}=n_{HI}/n$ (top) or $\rm 2x_{H_2}=2n_{H_2}/n$ (bottom) versus $\rm n$, where $\rm n_{HI}$ and $\rm n_{H_2}$ are the volume density of \ion{H}{1} and H$_2$ respectively. $\rm n$ is the total (atomic plus molecular) hydrogen volume density. P gives the volume fraction of each data point and the bin-size is 200.}
	\end{figure*}
	\begin{figure*}[t]
		\centering
		\includegraphics[width=1.00\linewidth,height=0.55\linewidth]{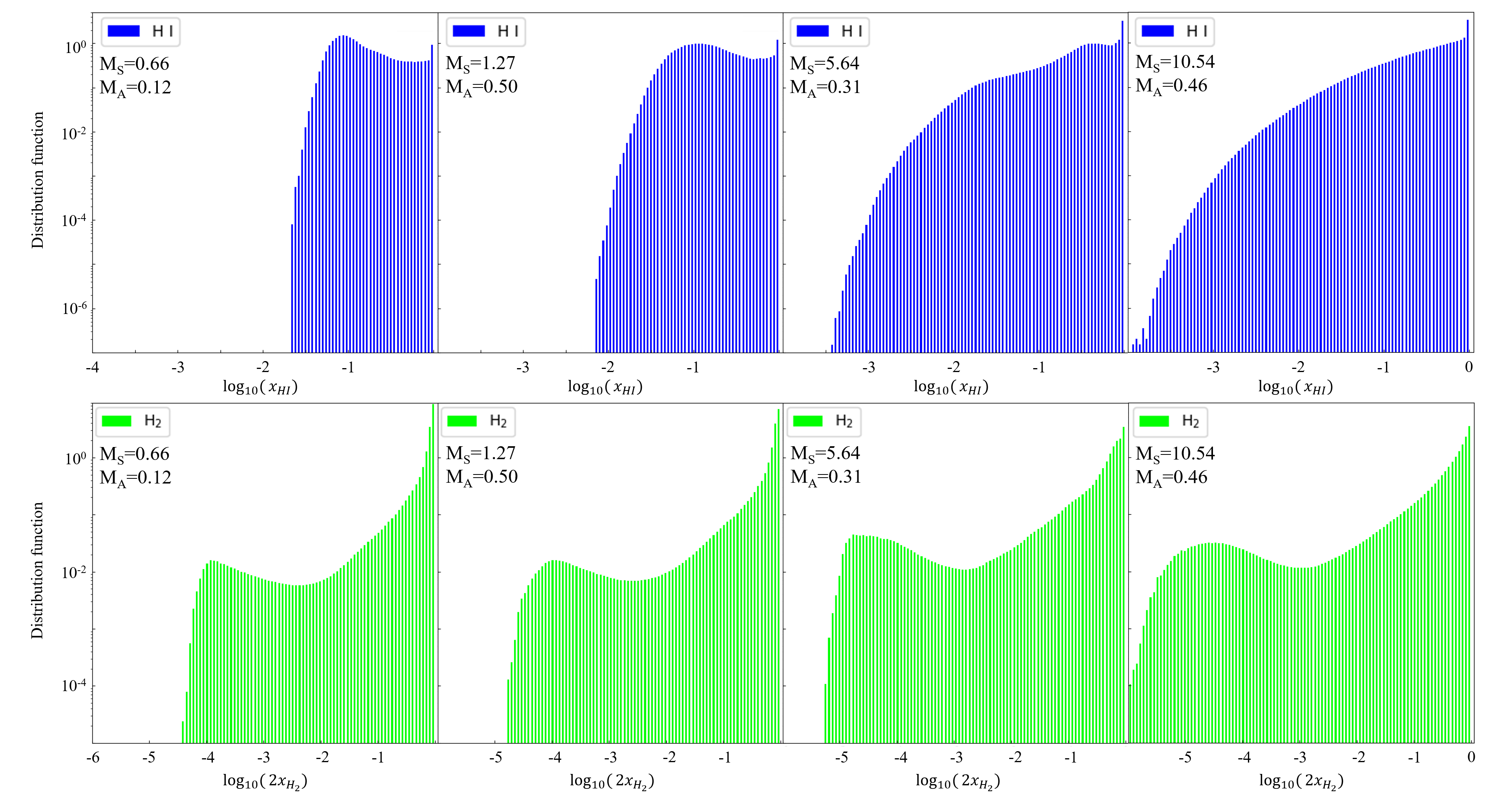}
		\caption{\label{fig:pdf} The normalized PDFs of $\rm x_{HI}=n_{HI}/n$ (blue) and $\rm 2x_{H_2}=2n_{H_2}/n$ (green). $\rm n_{HI}$ and $\rm n_{H_2}$ are the volume density of \ion{H}{1} and H$_2$ respectively. $\rm n$ is the total (atomic plus molecular) hydrogen volume density. The bin-size of the PDFs is 50.}
	\end{figure*}
	The PCA uses an orthogonal linear transformation to convert a set of possibly correlated variables to a set of linearly independent variables called principal components. For the implementation of PCA, the PPV cube $\rho(x,y,v)$ is assumed as a probability density function of three random variables $(x,y,v)$. We firstly calculate the covariance matrix and its corresponding eigenvalue equation \citep{2002ApJ...566..276B,2002ApJ...566..289B,PCA}:
	\begin{equation}
		\label{eq:13}
		\begin{aligned}
			M(v_i,v_j) \propto &\int dxdy \rho(x,y,v_i)\rho(x,y,v_j)\\
			&-\int dxdy \rho(x,y,v_i)\int dxdy\rho(x,y,v_j)
		\end{aligned}
	\end{equation}
	\begin{equation}
		\label{eq:14}
		\textbf{M}\cdot \textbf{u}=\lambda\textbf{u}
	\end{equation}
	where \textbf{M} is the co-variance matrix with matrix element $S(v_i,v_j)$, with $i,j=1,2,...,n_v$. $n_v$ is the number of channel in PPV cubes and $\lambda$ is the eigenvalues associated with the eigenvector $\textbf{u}$. The eigenvalues correspond to the weight of each principal component. If the eigenvalue is small, then the contribution from its corresponding principal component are also small. We can therefore enhance the contribution from crucial components by projecting the original data set into the new orthogonal basis formed by the eigenvectors \citep{PCA,EB}. The projection of the PPV cube into the new orthogonal basis is operated by weighting channel $\rho(x,y,v_j)$ with the corresponding eigenvector element $u_{ij}$, in which the corresponding eigen-channel $I_i(x,y)$ is:
	\begin{equation}
		I_i(x,y)=\sum_j^{n_v}u_{ij}\cdot \rho(x,y,v_j)
	\end{equation}
	Through this projection, we have totally $n_v$ eigen-channels in the PCA space. The recipe of gradient's calculation and the sub-block averaging method (see \S~\ref{sec:method}) are applied to each eigen-channel, so that we have the eigen-gradient fields $\psi_{gi}^s(x,y)$ with $i=1,2,...,n_v$. In analogy to the Stokes parameters of polarization, the pseudo Q$_g$ and U$_g$ of gradient-inferred magnetic fields are defined as:
	\begin{equation}
		\label{eq.17}
		\begin{aligned}
			& Q_g(x,y)=\sum_{i=1}^n I_i(x,y)\cos(2\psi_{gi}^s(x,y))\\
			& U_g(x,y)=\sum_{i=1}^n I_i(x,y)\sin(2\psi_{gi}^s(x,y))\\
			& \psi_g=\frac{1}{2}\tan^{-1}(\frac{U_g}{Q_g})
		\end{aligned}
	\end{equation}
	The pseudo polarization angle $\psi_g$ is then defined correspondingly, which gives a probe of plane-of-the-sky magnetic field orientation after rotating 90$^\circ$. Note in constructing the $Q_g(x, y)$ and $U_g(x, y)$, the number of used eigen-channel $n$ can be less than $n_v$. In \citet{PCA}, it was shown that only the first twenty eigen-channels are most important, and other eigen-channels are negligible and more sensitive to noise. Nevertheless, we still consider all eigen-channels when constructing the pseudo-Stokes parameter in this work. We use the abbreviation VGT+PCA to represent this method (see \citealt{EB} for the full recipe).
	
	\subsection{Alignment measure}
	To quantify the relative alignment between the magnetic fields and rotated gradient angle, we utilize the \textbf{Alignment Measure} (AM): 
	\begin{align}
		\label{AM_measure}
		\rm AM=2(\langle \cos^{2} \theta_{r}\rangle-\frac{1}{2})
	\end{align}
	where $\theta_r$ is the angular difference of two vectors, while $\langle ...\rangle$ denotes the average within a region of interests. In the case of a perfect alignment of the magnetic field and gradient, we get AM = 1, i.e., we have a perfect global alignment between gradients and the POS magnetic fields. AM = -1 indicates global gradients are perpendicular to the POS magnetic field and AM = 0 means globally random correlation. The standard error of the mean gives the uncertainty $\sigma_{AM}$; that is, the standard deviation divided by the square root of the sample size N:
	\begin{equation}
		\sigma_{AM}^2=\frac{\langle \cos^22\theta_r\rangle-\langle \cos2\theta_r\rangle^2}{N}
	\end{equation}
	The AM given by Eq. (\ref{AM_measure}) was introduced in \citet{2017ApJ...835...41G} and later was borrowed by other researchers doing their studies, e.g. the studies of the Rolling Hough Transform \citep{2019ApJ...887..136C}. Similarly, \citet{2018MNRAS.474.1018J} introduce the projected Rayleigh statistic (PRS) for characterising
	relative orientations. For a set of N 
	angles $\theta_i \in [-\pi/2, \pi/2]$, the PRS $Z_x$ and its uncertainty $\sigma_{Z_x}$ are defined as: 
	\begin{equation}
		\begin{aligned}
			Z_x&=\frac{\sum_i^N\cos 2\theta_i}{\sqrt{N/2}}\\
			\sigma^2_{Z_x}&=\frac{2\sum_i^N(\cos 2\theta_i)^2}{N}-\frac{2(\sum_i^N\cos 2\theta_i)^2}{N^2}\\
		\end{aligned}
	\end{equation}
	therefore, $Z_x\gg0$ indicates strong parallel alignment, while $Z_x\ll0$ indicates strong perpendicular alignment. When the sampling size N is large. Comparing with PRS, the AM takes the average directly so that its value is bounded between [-1, 1]. In this work, we only use AM to characterize the relative orientation.
	
	\begin{figure}[t]
		\centering
		\includegraphics[width=1.0\linewidth,height=1.32\linewidth]{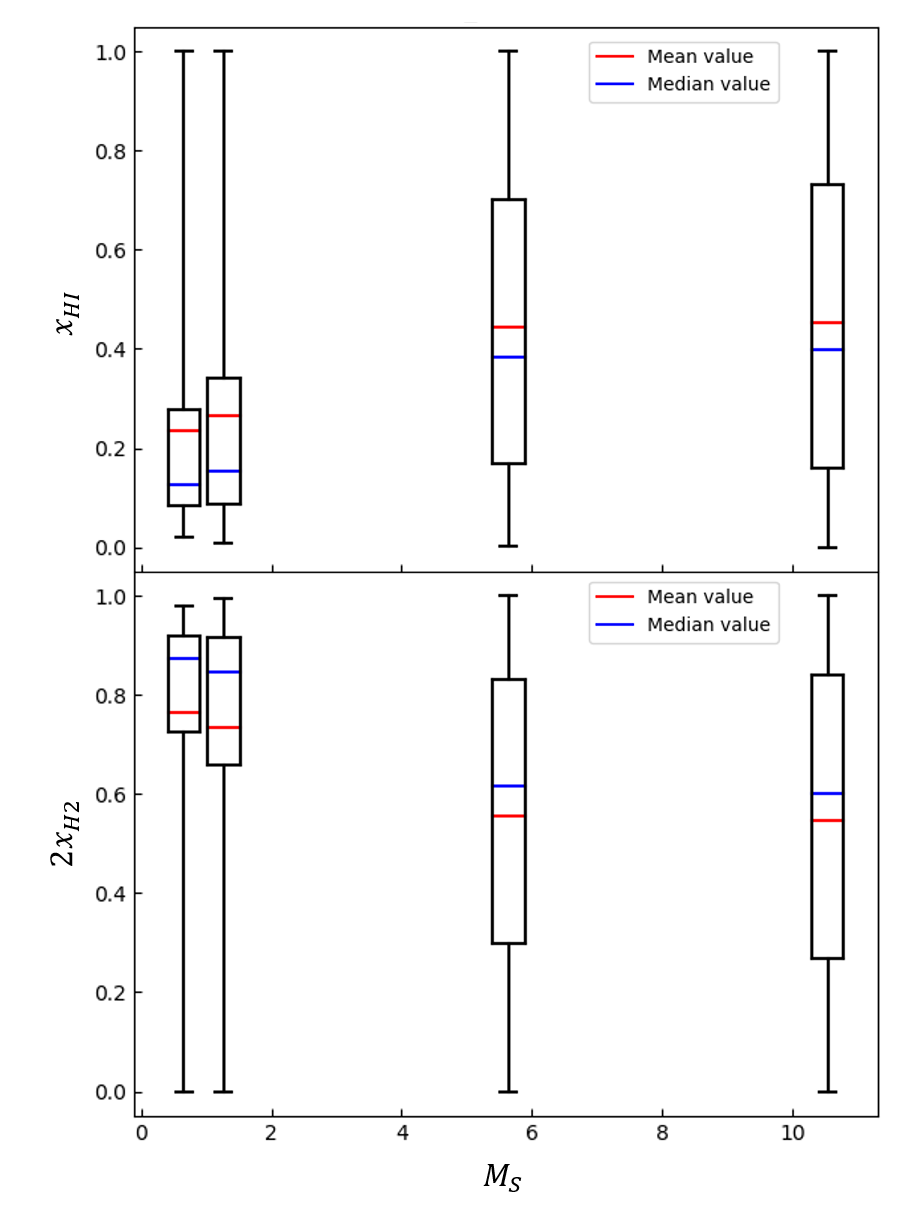}
		\caption{\label{fig:fra} The statistical properties of $\rm x_{HI}$ and $\rm 2x_{H_2}$ distribution. The line along y-axis give the range of $\rm x_{HI}$ and $\rm 2x_{H_2}$. The box indicates the interquartile range and the red/blue segment represents the mean/median value. The entire vertical segment gives the range of $\rm x_{HI}$ and $\rm 2x_{H_2}$}.
	\end{figure}

	\section{Result}
	\label{sec:result}
	\subsection{The distributions of the \ion{H}{1} and H$_2$ abundances}
	
	In this section, we discuss the distributions of the atomic and molecular fractions, $\rm x_{HI}$ and $\rm x_{H2}$, for nonhomogeneous turbulent gas. These distributions are shown in a set of three figures, Figs.~\ref{fig:hist}, \ref{fig:pdf}, and \ref{fig:fra}. In these figures we show the 2D PDFs of $n$ versus $\rm x_{HI}$ and $\rm x_{H2}$, standard 1D PDFs of  $\rm x_{HI}$ and $\rm x_{H2}$, and a bar plot showing a statistical summary of  the atomic and molecular fractions.
	
	Examining the four panels in Fig.~\ref{fig:hist}, from left to right, we see that at small $\rm M_S$, the distribution is narrow in the $y$ direction, i.e., the dynamic range in the density is small, as expected for subsonic gas. With increasing $\rm M_s$, density fluctuations develop in the gas, and the density dispersion increases. At a given density, the dispersion of the \ion{H}{1} and H$_2$ PDFs along the $x$-axis results from the transition from atomic ($\rm x_{HI}=1$) to molecular state ($\rm 2x_{H2}=1$). The dispersion increases with cloud depth. In particular, the low values of $\rm 2x_{H2}$ (and high values of $\rm x_{HI}$) correspond to regions near the cloud boundary where H$_2$ destruction is very efficient. With increasing column density (i.e., distance from cloud edge), the UV radiation is absorbed by dust and in the H$_2$ lines (H$_2$ self-shielding), the H$_2$ photodissociation rate decreases and $x_{H2}$ increases. Finally, at sufficient cloud depth, the gas becomes molecular. The \ion{H}{1} fraction never falls to zero as even in cloud interiors cosmic rays maintain a small H population \citep{2003ApJ...585..823L}.
	
	The balance between \ion{H}{1} formation via cosmic-rays in cloud interiors, and its destruction (to form H$_2$) gives rise to the diagonal shape seen at the lower end of the \ion{H}{1} PDFs in Fig.~\ref{fig:hist}. Following Eq.~\ref{eq: H-H2 balance} at large cloud depth the photodissociation term is negligible, $\rm 2x_{H2}\simeq 1$ and we get $\rm x_{HI}\propto 1/n$. Similarly, in all the panels, the diagonal shape of the H$_2$ PDFs at their low end results from the balance of un-shielded H$_2$ photodissociation and H$_2$ formation at cloud edge. This results in $\rm x_{H_2} \propto n$ (i.e., Eq.~\ref{eq: H-H2 balance} with $\zeta \ll D_0$ and with the shielding terms $\approx 1$). For a complementary discussion of 2D PDFs of density and molecular abundances see \citet{2019ApJ...885..109B}.
	
	Throughout the box, the H$_2$ formation rate (= the \ion{H}{1} destruction rates) increase with $n$. Thus, positive fluctuations in the density promote higher H$_2$ fractions and lower \ion{H}{1}, resulting in the observed correlation between $\rm x_{H_2}$ and $n$. In cloud interiors, the dependence is not linear as besides the H$_2$ formation rate (which is linear in $n$), the destruction of H$_2$ is strongly affected by H$_2$ self-shielding, which in turn depends on the accumulated H$_2$ column ($\rm N_{H2}=\int n x_{H2} dz$), from cloud edge to the point of interest \citep[][]{2017ApJ...843...92B}. In addition, we observe there exists an artificial tail in the low values of $\rm 2x_{H2}$ (or high values of $\rm x_{HI}$). This tail becomes more apparent in super-sonic cases. This results from the finite resolution of the grid, and the tail corresponds to the $\rm x_{H2}$ at the cloud edge (see Fig.~3 in \citealt{2019ApJ...885..109B} and discussion in \S~4.1). The transition occurs over very short length-scales at high densities, which correspond to one unit grid or less. When examining the 2D plot from left to right at constant density, the $\rm 2x_{H2}$ jumps from the value at the tail (cloud edge) to its maximum value $\rm 2x_{H2}\approx1$ in the cloud interior. At lower density, the transition is continuous, as many cells resolve the transition. That is also why there is a large volume fraction in the low values of $\rm 2x_{H2}$. 
	
	Fig.~\ref{fig:pdf} shows 1D PDFs of the H and H$_2$ abundances.
	Here, again we see how the \ion{H}{1} and H$_2$ PDFs broaden with increasing sonic Mach number. The two peaks correspond to the cloud-edge and inner cloud regions. The region between the two peaks results mainly from intermediate cloud depth, where the transition from H to H$_2$ takes place. At large sonic Mach numbers, the lower end of the H$_2$ PDF broadens due to density fluctuations. Without them, the H$_2$ abundance at the cloud edge is single-valued. With density fluctuations, $\rm x_{H2}\propto n$ at the cloud edge, and the low-end of the PDF then reflects the density PDF, which is nearly lognormal. The effect of density fluctuations is also seen in the \ion{H}{1} PDF, which broadens to smaller values with increasing $\rm M_S$. This tail results from regions at deep cloud interiors where \ion{H}{1} forms via the action of cosmic-rays and is destroyed by the H$_2$ formation process. At large cloud interiors this results in $\rm x_{HI} \propto 1/n$ and thus high density regions push the \ion{H}{1} fraction to lower values.
	
	In Fig.~\ref{fig:fra}, we show the interquartile range, mean value, and median value of the $\rm x_{HI}$ and $\rm x_{H_2}$ distributions. As in the previous figures, again, we see the trend of an increasing dispersion with increasing $\rm M_s$ as strong density fluctuations develop in the gas. In addition, we see that with increasing $\rm M_s$, the median and mean values of $\rm x_{H2}$ decrease, while that of \ion{H}{1} increases. The medians' trend may be understood in terms of the properties of the density ($n$) PDF in turbulent gas. Let $x \equiv n/\langle n \rangle$. For non-self-gravitating, magnetized, and isothermal turbulence $x$ follows a lognormal distribution \citep{1995ApJ...441..702V,2008ApJ...680.1083R,2012ApJ...750...13C,2018ApJ...863..118B}, for which $x_{\rm median}<1$ and $x_{\rm median}$ decreases with increasing $\rm M_s$. Physically, with increasing sonic Mach number, larger regions of the volume are filled with moderately low-density material, $x<1$, compensated by small volumes filled with very high $x$; i.e., shocks. The abundant low-density regions result in lowered H$_2$ abundances and increased H, resulting in the trend seen in Fig.~\ref{fig:fra}. The means' behavior is more involved as the H, and H $ _2$ abundances depend non-linearly (and non locally) on the gas density.

	\begin{figure*}[p]
		\centering
		\includegraphics[width=0.90\linewidth,height=1.25\linewidth]{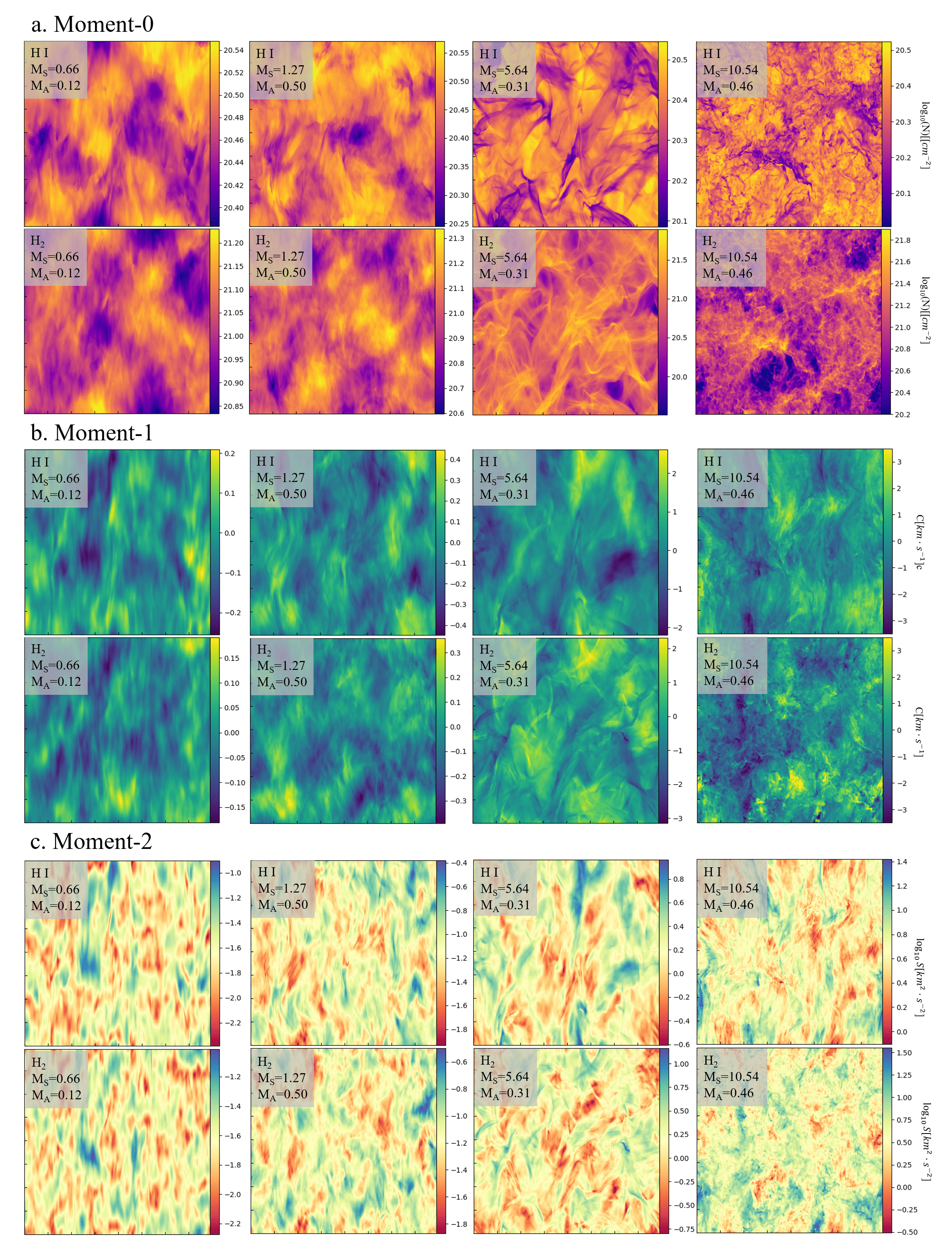}
		\caption{\label{fig:maps} The moment-0 maps (panel a), moment-1 maps (panel b), and moment-2 maps (panel c) for each simulation listed in Tab.~\ref{tab:sim}. For each panel, the \ion{H}{1} maps are in the top row, which H$_2$ maps are in the bottom row.}
	\end{figure*}
	
	\begin{figure*}[p]
		\centering
		\includegraphics[width=0.87\linewidth,height=1.3\linewidth]{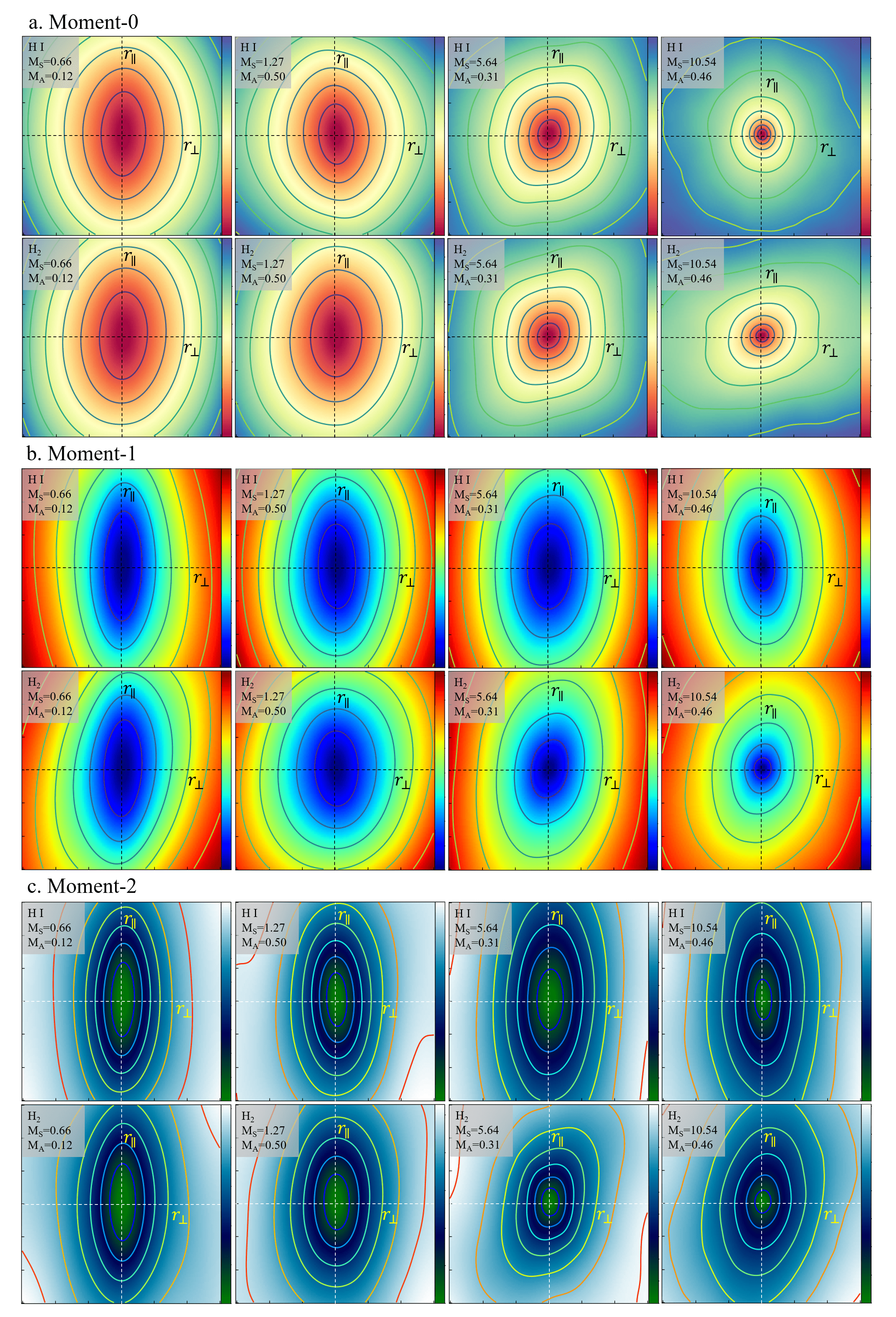}
		\caption{\label{fig:SF} The structure function of moment-0 maps (panel a), moment-1 maps (panel b), and moment-2 maps (panel c) for each simulation listed in Tab.~\ref{tab:sim}. For each panel, the \ion{H}{1} maps are in the top row, which H$_2$ maps are in the bottom row. $r_\bot$ and $r_\parallel$ are the real space scales perpendicular and parallel to the magnetic field respectively. For all plots, $r_\bot$ and $r_\parallel$ are in scales less than 60 pixels.}
	\end{figure*}
	
	\begin{figure*}[t]
		\centering
		\includegraphics[width=1.0\linewidth,height=0.85\linewidth]{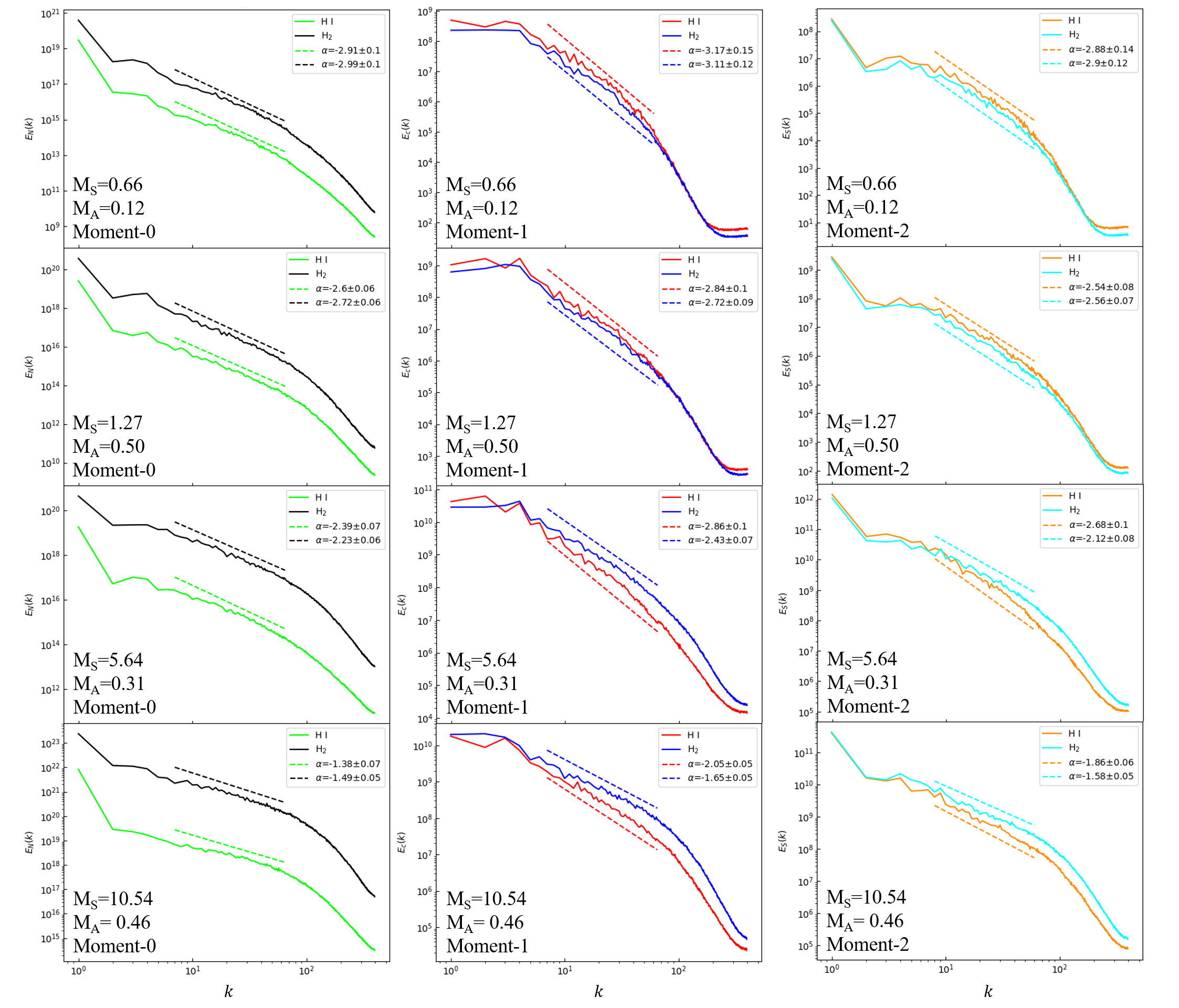}
		\caption{\label{fig:spectrum} The energy spectrum of the moment-0 maps (the 1st column), moment-1 maps (the 2nd column), and moment-2 maps (the 3rd column) for each simulation listed in Tab.~\ref{tab:sim}. $\alpha$ is the fitted slope of the spectrum.}
	\end{figure*}
	
	\subsection{Moment maps of \ion{H}{1} and H$_2$ gas}
	The moment maps can carry the density information and velocity information of the turbulent medium. For example, the moment-0 map contains a pure density field, while moment-1 and moment-2 maps mix both density and velocity field (see \S~\ref{sec:theory}). The moment-maps shall exhibit similar anisotropic properties as the density and velocity field, i.e., elongating along with their local magnetic fields. In PDRs, however, the density field is likely changed due to the formation of other molecules. This transition process also introduces velocity fluctuations. Nevertheless, the velocities induced by H$_2$ formation is not turbulent, as they do not form any eddies. These velocities are differences of thermal velocities that disappear over just one mean free path length. Therefore, the velocity field is expected to be anisotropic still. In this section, we study how the transition process affects the anisotropy in moment-maps.
	
	In Fig.~\ref{fig:maps}, we present the moment-0, moment-1, and moment-2 maps for each simulation. For sub-sonic moment-0 maps, we can see the intensity structures are elongating along the vertical direction, i.e., the mean magnetic field's direction. However, for the super-sonic turbulence M$_s>$1, the anisotropic pattern becomes less significant due to shocks. As explained in \citet{2019ApJ...878..157X}, these dense shock structures are preferentially perpendicular to the magnetic field. Therefore, in super-sonic turbulence, we can observe the dense structures being perpendicular to the magnetic field. However, in the process of \ion{H}{1}-to-H$_2$ transition, the dense gas usually is sampled into H$_2$ so that H$_2$ map contains more shocks then \ion{H}{1} map. We, therefore, can expect the H$_2$ map is less anisotropic. 
	
	The anisotropic pattern that the structures elongate along the mean magnetic field's direction can also be observed in moment-1 and moment-2 maps. The difference here is that moment-1 and moment-2 maps contain the velocity contribution, which is insensitive to shocks (see \S~\ref{sec:theory}). The effect of shocks is hence diluted and the moment-1/moment-2 maps can keep anisotropic even in super-sonic cases.
	\begin{figure*}[t]
		\centering
		\includegraphics[width=1.00\linewidth,height=0.6\linewidth]{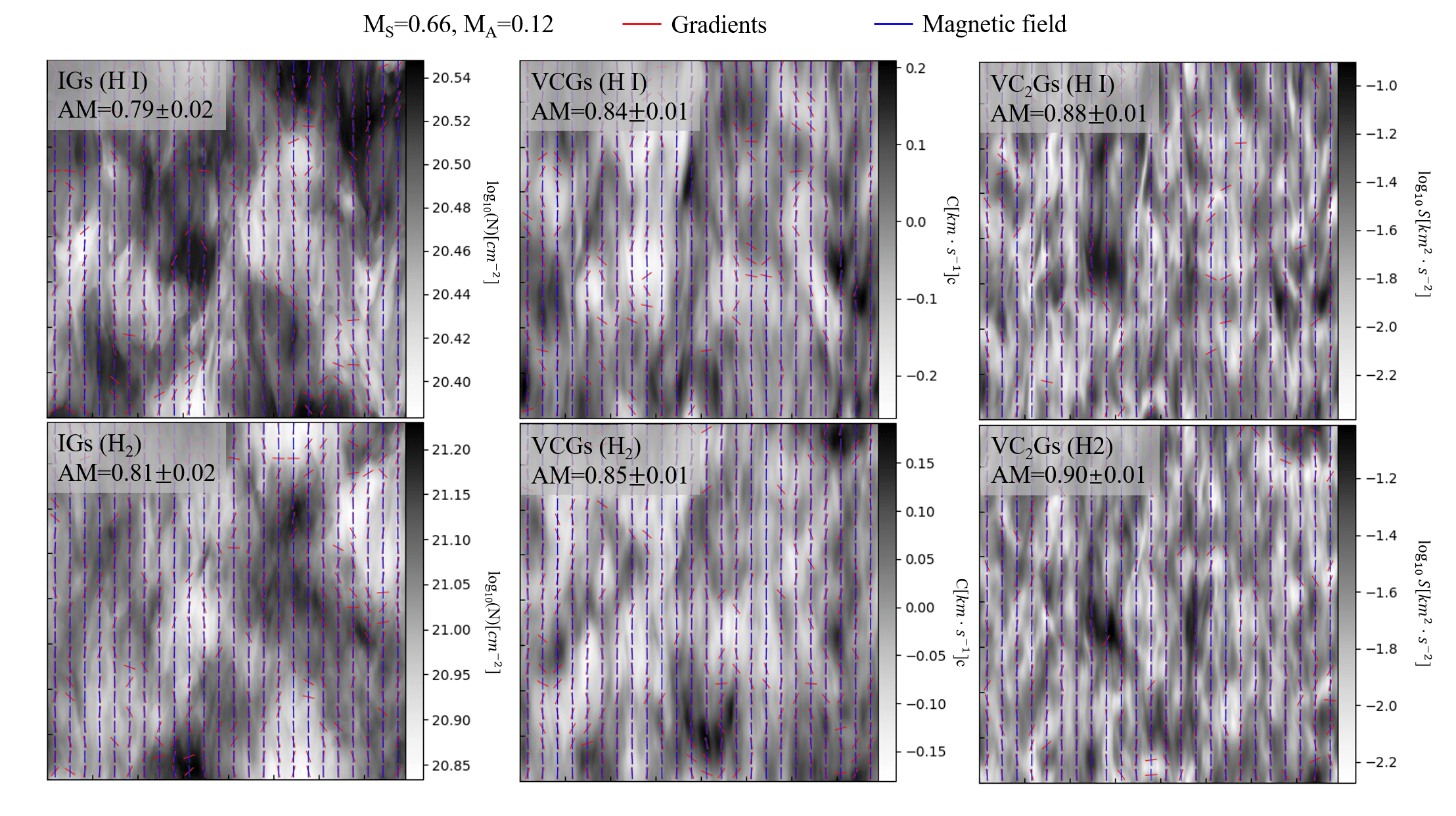}
		\caption{\label{fig:gradient} Plot of the magnetic field orientation probed by GT (red) and synthetic dust polarization (blue). The plot is overlaid on the moment-0 maps (left), moment-1 maps (middle), and moment-2 maps (right). We use the \ion{H}{1} (top) and H$_2$ (bottom) cubes produced by simulation A0 and denote the gradient of moment-0 map as IGs, moment-1 map as VCGs, and moment-2 map as VC$_2$Gs.}
	\end{figure*}
	\begin{figure*}[t]
		\centering
		\includegraphics[width=1.00\linewidth,height=0.53\linewidth]{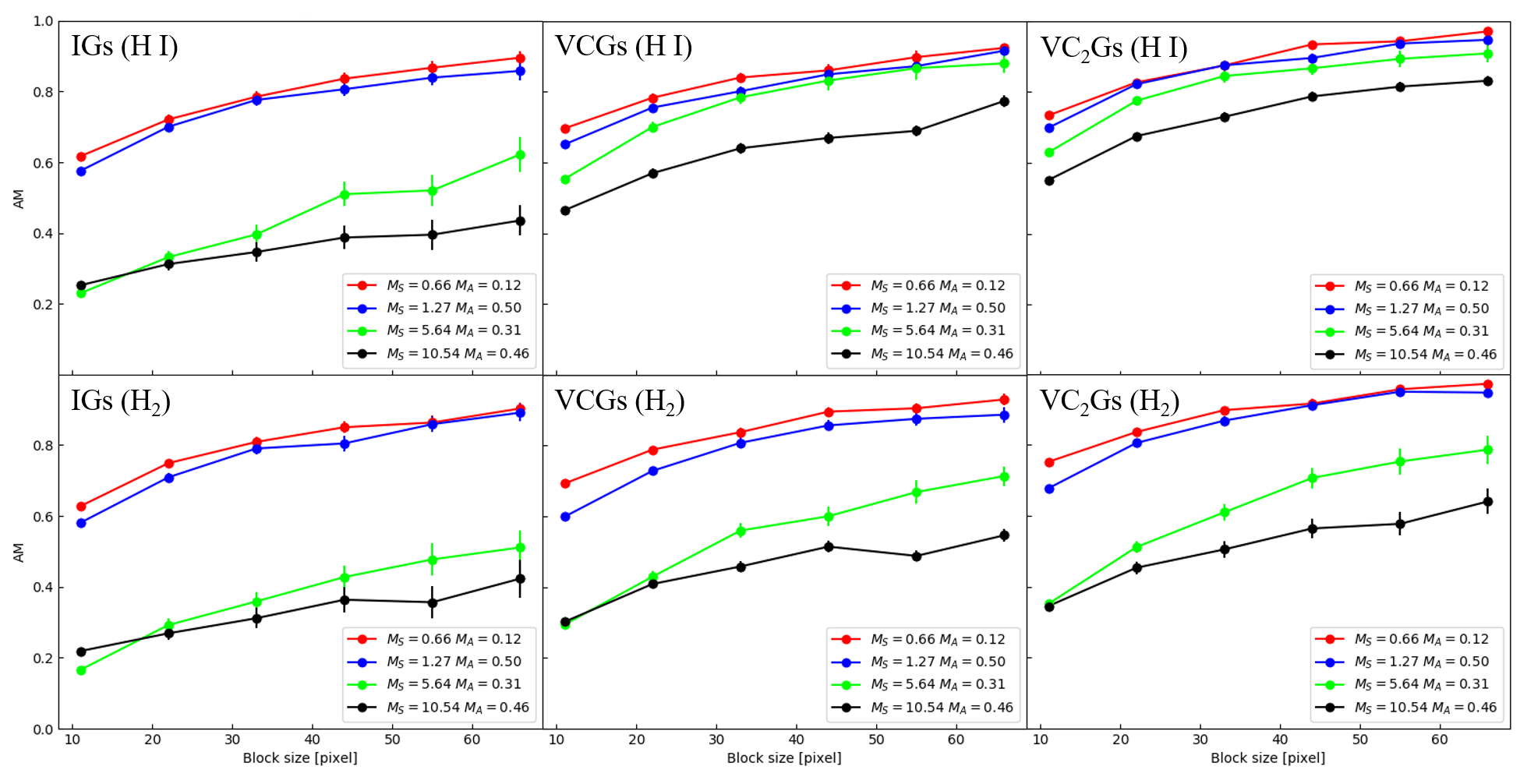}
		\caption{\label{fig:am} The correlation between AM and the sub-block size, which is selected for the implementation of the sub-block averaging method. We denote the gradient of moment-0 map as IGs, moment-1 map as VCGs, and moment-2 map as VC$_2$Gs.}
	\end{figure*}
	\begin{figure*}[t]
		\centering
		\includegraphics[width=1.00\linewidth,height=0.45\linewidth]{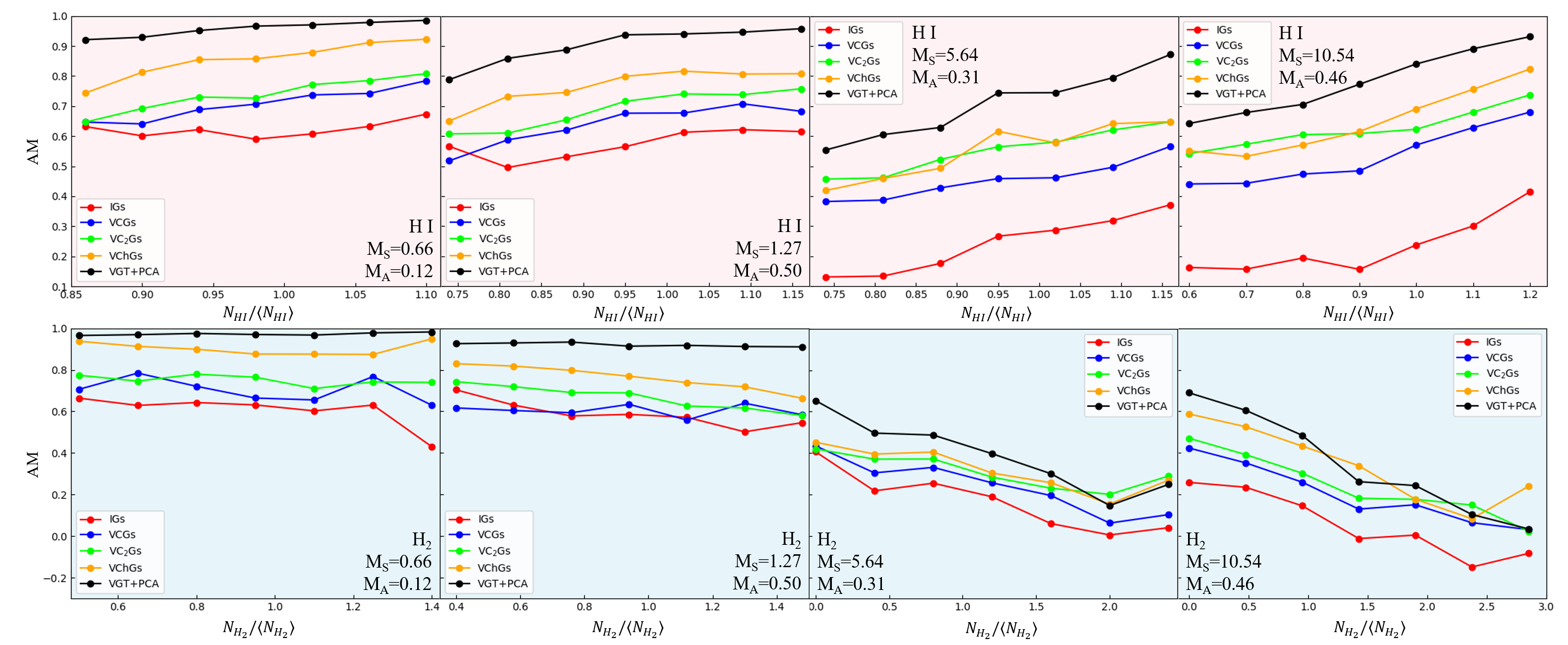}
		\caption{\label{fig:zscore} The correlation between AM and the ratio of $\rm N_{HI}/\langle N_{HI}\rangle$ (top, red background) and $\rm N_{H_2}/\langle N_{H_2}\rangle$ (bottom, blue background). $\rm N_{HI}$/$\rm N_{H_2}$ is the column density of \ion{H}{1}/$\rm H_2$ and $\rm \langle...\rangle$ denotes the mean value.}
	\end{figure*}
	
	\subsubsection{Characterise the anisotropy by structure function}
	\label{subsec:sf}
	The structure function is widely used to characterise the anisotropic properties of MHD turbulence. Here, we apply the second-order structure function to the moment maps, which is defined as:
	\begin{equation}
		SF(\Vec{R})=\langle [f(\Vec{r})-f(\Vec{r}+\Vec{R})]^2\rangle
	\end{equation}
	here $\Vec{r}=(x,y)$, $\Vec{R}$ is a lag vector, and $f(\Vec{r})$ can be replaced by I($\Vec{r}$), C($\Vec{r}$), or S($\Vec{r}$). In terms of the structure-function, the anisotropy appears on a small scale. We therefore typically constrain $\Vec{r}$ in the range of fewer than 60 pixels. 
	Fig.~\ref{fig:SF} shows how the structure-function should behave in terms of the contour plot. We use $r_\bot$ and $r_\parallel$ to denote the real space scales perpendicular and parallel to the mean magnetic field, respectively.
	
	For the structure-function of moment-0 maps,  we can observe that in the case of sub-sonic turbulence, the contours of both \ion{H}{1} and H$_2$ are elongated along the $r_\parallel$ direction, i.e., the mean magnetic field direction. The contours on a large scale are slightly misaligned from the $r_\parallel$ direction because of the limited inertial range in numerical simulations \citep{2018ApJ...865...54Y}. For the trans-sonic turbulence $\rm M_S=1.27$, as the turbulence's compression starts its dominance, the large scale contours start evolving to approximately isotropic, while the small scale contours are still anisotropic. This phenomenon is more significant for the structure-function of \ion{H}{1} supersonic moment-0 maps. For H$_2$, we can still see the change of the structure functions. The contour is getting isotropic when $\rm M_s=5.64$, while it elongates in the direction perpendicular to the magnetic field (vertical direction) at large scales when $\rm M_S=10.54$. \citet{2020MNRAS.492..668B} also reports a similar finding, which is in agreement with theoretical expectations in \citet{2019ApJ...878..157X}. The authors numerically demonstrated that for sub-Alfv\'{e}n turbulence with $\rm M_s<4$, the anisotropy in the column density is dominated by striations aligned with the magnetic field. However, \citet{2020ApJ...894L...2B} find 3D density structures to squeeze along the magnetic fields in moderately supersonic gas. The anisotropy is significantly changed by high-density structures that form perpendicular to the magnetic field for sub-Alfv\'{e}n and highly supersonic turbulence. We observe that the moment-1 and moment-2 maps have different behavior. For moment-1 maps, the anisotropy is always parallel to the magnetic field. Different from the structure-function of \ion{H}{1}'s supersonic moment-0 maps, the structure functions of moment-1 and moment-2 maps are striated in the direction parallel to the magnetic field (vertical direction). In addition,  the structure-function of super-sonic \ion{H}{1} is more striated (i.e., a larger semi-major-axis to semi-minor-axis ratio) than the one of H$_2$. The shock effect in H$_2$ is likely significant, although  moment-1 and moment-2 map partially dilutes shock contribution. In any case, the anisotropic patterns of moment-1 and moment-2 maps are always aligned with the magnetic field. This supports one of the arguments at the core of the GT, namely, that the velocity fluctuations are more reliable tracers of the magnetic field than density fluctuations, i.e., compared to the studies of \citet{Soler2013,2014ApJ...789...82C,2019ApJ...887..136C}.

	\subsubsection{Energy spectrum for each moment map}
	Furthermore, we calculate the energy spectrum for each moment maps. The results are presented in Fig.~\ref{fig:spectrum}. For moment-0 maps, dense H$_2$ contains more energy than \ion{H}{1}. This difference is more significant in super-sonic cases, as the majority of high-density contrasts induced by density fluctuation have been sampled into H$_2$. As for moment-1 and moment-2 maps, \ion{H}{1} and H$_2$ occupy approximately equivalent fraction of energy in sub/trans-sonic turbulence. Also, similarly, H$_2$ gets more energy in super-sonic moment-1 and moment-2 maps, which means dense and small-scale structures are sampled into H$_2$. 
	
	We fit the slope $\alpha$ of the spectrum in the range of $8\le k\le60$. We find the slope gets shallower with the increment of $\rm M_s$. In sub-sonic cases, both \ion{H}{1} and H$_2$ fits similar spectrum slopes for corresponding moment maps (moment-0 map: $\sim -2.9$, moment-1 map: $\sim -3.1$, and moment-2 map: $\sim-2.9$). With the increment of $\rm M_s$, the spectrum of H$_2$ moment-1 and moment-2 maps becomes shallower the one of \ion{H}{1} moment-1 and moment-2 maps. However, the spectral of \ion{H}{1} and H$_2$ moment-0 maps always exhibit a similar slope value. The shallower spectrum can be caused by shocks, which was numerically demonstrated by \citet{2005ApJ...624L..93B}. For instance, moment-0 maps are produced by integrating spectral lines in full range (see Fig.~\ref{fig:los}). The contribution from density and all shocks in super-sonic turbulence is more significant in moment-0 maps. We, therefore, see a shallower slope in the energy spectrum. Also, the density fluctuation induced by super-sonic turbulence produces a number of small-scale structures, as shown in Fig.~\ref{fig:proj}. These small-scale structures also shallower the spectrum.
	
	\subsection{Tracing the magnetic field using GT}
	In the above, we discussed the anisotropy of atomic \ion{H}{1} and molecular H$_2$ species, which is the theoretical foundation of GT. We can see both \ion{H}{1} and H$_2$ maintain the anisotropic property after the transition. In this section, we apply GT to trace the magnetic field and compare its performance with the utilization of different moment maps. Note we denote the gradient of moment-0 map, or intensity map, as IGs, moment-1 map, or first-order centroids map, as VCGs, and moment-2 map, or second-order centroids map, as VC$_2$Gs.
	
	In Fig.~\ref{fig:gradient}, we give a comparison of the magnetic field orientation probed by synthetic dust polarization and by GT with a block size of 33 pixels. The plot is overlaid on the moment maps of \ion{H}{1} and H$_2$ cubes produced by sub-sonic simulation A0. Comparing the AM, we find the VC$_2$Gs have a better agreement with the magnetic field than VCGs and IGs. Also, owing to the fact that the compression of turbulence is insignificant in the absence of shocks, \ion{H}{1} and H$_2$ give similar anisotropic properties. Therefore, we can see \ion{H}{1} and H $ _2$ maps produce similar AM values, although H $ _2$ samples denser gas.
	
	We further study the alignment of gradients and the magnetic field in terms of tracing scale. In Fig.~\ref{fig:am}, we vary the sub-block size from 11 to 66 pixels and calculate the AM correspondingly. In general, the AM is positively proportional to the sub-block size since a larger sub-block guarantees more sample vectors so that the mean magnetic field inferred from the gradients is statistically better determined \citep{YL17a,LY18a}. Empirically, the minimum sub-block size for the analysis of synthetic simulations was found to be 20 pixels \citep{LY18a}, which provides sufficient 400 samples for the Gaussian fitting in the absence of noise. For the \ion{H}{1} gradients, the sub-sonic and trans-sonic cubes give similar results as the turbulence is dominating. As for super-sonic cubes, theoretically, high-density shocks flip the density gradient by 90$^\circ$ being parallel to the magnetic field \citep{YL17b,IGs}. Therefore, the alignment AM is anti-correlated with $\rm M_s$. The shocks' effect is more apparent for H$_2$ gradients, particularly for VCGs and VC$_2$Gs. It is likely that H$_2$ samples more shocks than \ion{H}{1} so that velocity contribution in VCGs and VC$_2$Gs can only dilute limited the shock effect. Nevertheless, we find the VC$_2$Gs of \ion{H}{1} is the most stable tracers of the magnetic field in the presence of shocks.
	
	Additionally, we examine our hypothesis about shocks. Firstly, we calculate the AM in each sub-block instead of taking global averaging. We then sort out the AM values based on the ratio between their corresponding column density and global mean column density, i.e., $\rm N_{H_I}/\langle N_{H_I}\rangle$ and $\rm N_{H_2}/\langle N_{H_2}\rangle$. The moment-0 map is segmented into seven groups based on the ratio. Note that as the range of $\rm N_{H_2}/\langle N_{H_2}\rangle$ and $\rm N_{H_I}/\langle N_{H_I}\rangle$ is not a fixed value but depends on the sonic Mach number, we do not keep a constant interval when doing segmentation for different maps. In each segmentation, we output the averaged AM value. In Fig.~\ref{fig:zscore}, we plot the AM value versus $\rm N_{H_2}/\langle N_{H_2}\rangle$ and $\rm N_{H_I}/\langle N_{H_I}\rangle$. The sub-block size is selected as 20 pixels, and the moment-0 map is also reduced to the same resolution before doing the segmentation. 
	
	For the sub-sonic \ion{H}{1} case, we see that the AM is slightly increasing when the ratio of $\rm N_{H_I}/\langle N_{H_I}\rangle$ is large. When $\rm M_S$ is larger than 1, the AM starts decreasing but is still positively correlated with the $\rm N_{H_I}/\langle N_{H_I}\rangle$ ratio. Similarly, for H$_2$ gradients, the AM is decreasing in super-sonic cases but is negatively correlated with the $\rm N_{H_2}/\langle N_{H_2}\rangle$ ratio. When $\rm N_{H_2}/\langle N_{H_2}\rangle$ is close to its maximum value, the AM of IGs gets a minimum value around 0.6, 0, and -0.2 for $\rm M_S=1.27$, $5.64$, and $10.54$ respectively. The rapid decrease of AM in super-sonic H$_2$ high-density regions confirms our theoretical consideration about shocks, i.e., shocks can flip the gradients' direction. This agrees with the results in \citep{IGs}. Also, in \S~\ref{sec:theory}, we derived that the IGs consists of the contribution from the density gradient, while VCGs and VC$_2$Gs employ both density gradient and velocity gradient. We can see VCGs, in general, gets more resistance for shocks than IGs due to the velocity contribution. As VC$_2$Gs contains one additional velocity term than VCGs, VC$_2$Gs usually have better performance than VCGs in both sub-sonic and super-sonic cases.
	
	Furthermore, in Fig.~\ref{fig:zscore}, we introduce the comparison of VChGs and VGT+PCA. We can see in the super-sonic case that the AM of VChGs decreases with the increment of $\rm N_{H_2}/\langle N_{H_2}\rangle$. This is expected as the thin channel criteria Eq.~\ref{eq1} only indicates the velocity fluctuation is \textbf{dominating} over the density fluctuation, instead of pure velocity fluctuation. As we discussed above, IGs contains only density contribution and is most sensitive to shocks. The additional velocity contribution in VCGs and VC$_2$Gs (see \S~\ref{sec:theory}) guarantees them a better performance (higher AM). Most importantly, in any case, VChGs is superior to IGs, VCGs, and VC$_2$Gs. It means that VChGs definitely is dominating by the velocity contribution. The velocity contribution in the thin velocity channel is even more than the one in the moment-2 map.
	
	Nevertheless, we also see the VGT+PCA gives the best performance in tracing the magnetic field. We expect three possible reasons. First of all, VChGs utilizes only the central channel. It is likely the central channel only contains part of the magnetic field's information. Differently, the gradient from VGT+PCA is summed along the LOS for all velocity channels, so that all information in the PPV cube is taken into account. Also, the vector summation used in IGs, VCGs, VC$_2$Gs, and VChGs is different from the polarimetry, which employs the summation of Stoke parameters. However, the pseudo-Stoke parameters used in VGT+PCA is more similar to the polarimetry measurement. Additionally, the PCA technique enhances the most crucial components for constructing the gradient field. Noise is suppressed in this case \citep{PCA}.

	\section{Discussion}
	\label{sec:diss}
	\subsection{The role of density fluctuation in \ion{H}{1}-to-H$_2$ transition}
	The conversion of atomic \ion{H}{1} to molecular H$_2$ is a basic process in the turbulent, magnetized, interstellar medium. This transition has been widely studied from various perspectives, including numerically and analytically absorbing the effects of turbulence \citep{1995ApJ...440..674X,2002ApJ...573L.119W,2017ApJ...843...92B}. In this paper, we generate synthetic \ion{H}{1} and H$_2$ cubes in chemical balance to study the turbulence’s properties of \ion{H}{1}-to-H$_2$ transition. We fix the mean volume density $\langle \rm n \rangle \approx 50 cm^{-3}$ and temperature $\rm T=50K$, but varying the sonic Mach number $\rm M_S$ from 0.66 to 10.54. We find that in the case of supersonic turbulence, the distribution of H$_2$ fraction $\rm x_{H_2}$ and \ion{H}{1} fraction $\rm x_{HI}$ is more dispersed than the one of sub-sonic turbulence.  The increased dispersion reflects the growth of density fluctuation in supersonic turbulence \citep{2017ApJ...843...92B, 2019ApJ...885..109B}. For supersonic turbulence, the density fluctuation produces significant high-density and low-density contrasts. The high-density contrasts are fully sampled into H$_2$ so that the fraction of \ion{H}{1} left is much lower so that $\rm x_{HI}$ is more dispersed. Similarly, low-density contrasts are not sufficient for self-shielding so that the fraction of H$_2$ left is lower. This suggests that density fluctuations induced by turbulence can play a critical role in the formation of heavy molecular species.
	
	\subsection{Tracing local magnetic fields using velocity gradients}
	In addition to turbulence, the magnetic fields are also essential in regulating molecules’ formation. With the assistance of polarized dust emission, the projected magnetic field along LOS can be accessed. However, it is still challenging to separate the foreground and probing the local magnetic fields in either PDR or the molecular cloud. 
	
	A possible solution is provided by the Gradients Technique (GT), which has been developed as an advanced synergetic tool for studying the magnetic field \citep{2017ApJ...835...41G,YL17a,LY18a, IGs}. This technique uses the fundamental properties of MHD turbulence, namely, it is a sum of anisotropic eddies aligned, revealing the direction of the magnetic field in their vicinity \citep{GS95,LV99}. Due to the turbulent eddies being elongated along with their local magnetic fields, the corresponding density and velocity gradients are perpendicular to the magnetic fields.
	One can therefore infer the magnetic field orientation by rotating the gradients with 90$^\circ$. 
	
	One important advantage of GT is that it can utilize multiple molecular emission lines to construct the local magnetic field model. For instance, by using $\rm ^{12}CO$, $\rm ^{13}CO$, and $\rm C^{18}O$ data, GT can probe the POS component of the magnetic field over three different volume density ranges, from $10^2$ to $10^4$ $\rm cm^{-3}$ \citep{velac,2019ApJ...873...16H}. A similar idea can be migrated to the PRD regions. Since the H$_2$ usually samples denser gas than \ion{H}{1}, GT can also be used to trace the magnetic field in either H$_2$ or \ion{H}{1} region. 
	
	The performance of GT has been tested in either pure atomic \ion{H}{1} gas \citep{YL17a,2019ApJ...874...25G,EB,2020RNAAS...4..105H,HL20b} or pure molecular species \citep{survey,velac,2019ApJ...873...16H}. However, for the PDRs at the cloud boundaries, the transition from atomic gas to molecular gas could distort the turbulence's anisotropic properties crucial for GT. Therefore, in this work, we examine the effect of turbulence on interstellar clouds' chemical structure and GT's performance in PDRs. We find the anisotropy of pure density structures may be distorted in the transition process. The distortion depends on the spatial distribution of the gas and may affect the performance of GT. However, the H$_2$ formation taken place in a turbulent flow induces the change of the mass of the particles in the flow, but does not change the momentum and energy of the turbulent motions. The velocities induced by H$_2$ formation are differences of thermal velocities that disappear over just one mean free path length. These velocities are not turbulent in this case and they do not form any eddies. Also, \citet{2017ApJ...843...92B} showed that the \ion{H}{1}-to-H$_2$ transition occurs as UV Lyman-Werner photons are absorbed in the H$_2$ line wings, which are far (in wavelength) from spectral line centers. The gas velocity shifts the line centers by a small amount compared to the spacing between the lines, thus barely affecting the \ion{H}{1}-to-H$_2$ transition. It is important only for small columns where the gas is predominantly atomic, i.e., the column density $N\simeq 10^{19}-10^{20}$ cm$^{-2}$. Therefore, the turbulent velocity field is expected to be anisotropic and the velocity gradient is still a reliable tracer. 
	
	\subsection{Gradients of moment maps}
	To calculate the density or velocity gradients, GT usually employs the gradients of intensity map (i.e., moment-0 map, \citealt{YL17b,IGs}), velocity centroid map (i.e., moment-1 map, \citealt{2017ApJ...835...41G, YL17a}), and thin velocity channel map \citep{LY18a}. In this work, we introduce the second-order centroid map calculated from the PPV cube. We derive and confirm that the 2D gradient field of the intensity map carries only the information of turbulent density, while the 2D gradients of the velocity centroids contain the contribution from both the density and velocity fields. We find that the velocity contribution improves the performance of GT in tracing the magnetic field. This corresponds to theoretical expectations. If the density fluctuations are associated with the entropy variations, they can be passively transformed by the anisotropic velocity cascade and, therefore, reflect the velocity scaling properties. This can happen in subsonic turbulence, while in supersonic MHD turbulence, the properties of density and velocity can be very different \citep{2005ApJ...624L..93B,2007ApJ...658..423K}. In this situation, the magnetic field tracing with intensity gradients is not reliable because of shocks' presence. 
	
	According to the theory in \citet{LP00}, velocity fluctuation can also be obtained from a thin velocity channel maps. This has been developed as a branch of GT for studying the magnetic field \citep{LY18a}. However, the issue of what is measured in \ion{H}{1} channel maps has been debated recently. Some researchers believe that the striations observed in such maps are actual density filaments \citep{2014ApJ...789...82C}. The density, as we discussed earlier, for the velocity field can passively carry subsonic MHD turbulence. Therefore the density filaments can be aligned parallel to the magnetic field \citep{2019ApJ...878..157X}. However, it is also well known from the theory  \citep{LP00} and numerically demonstrated by different groups (see, \citealt{2018MNRAS.479.1722C,2018arXiv180200024Y}) that turbulent velocities produce the filaments in channel maps. Therefore, the pure density filament model is not tenable, and the actual question is the relative importance of turbulent densities and velocities for the observed striation. We note that by challenging the turbulent velocity origin of filaments, \cite{2019ApJ...874..171C} provided a series of arguments, the validity of which was seriously questioned in \citet{2019arXiv190403173Y}. Although several authors also argue that thin velocity channel maps observed in \ion{H}{1} emission are dominated by CNM \citep{2020ApJ...899...15M,2020arXiv200301454K}, this issue will be discussed in detail elsewhere.
	
	\subsection{Thermal Phases of the ISM}
	In this work, we use isothermal MHD simulations to study the chemical balance of \ion{H}{1}-to-H$_2$ transition. The isothermal \ion{H}{1}-to-H$_2$ transition model was proposed by \citet{2014ApJ...790...10S} and \citet{Bialy2016_tran} in the absence of turbulence. Analytically and numerically, \citet[][]{2017ApJ...843...92B, 2019ApJ...885..109B}, as well this work, extend the transition into an isothermal turbo-chemical model. The isothermal assumption (T = 50K)
	mimics a single-phase CNM. The cooling/heating processes in a real scenario, however, introduce a thermal instability resulting in a multi-phase medium, including both the CNM and the warm neutral media (WNM, \citealt{1969ApJ...155L.149F,1995ApJ...443..152W,2003ApJ...587..278W,2019ApJ...881..160B,2020arXiv200905466B}). In addition to the isothermal model, several authors also numerically considered the heating-cooling processes and found a bimodal PDFs of volume density, with a non-negligible mass within the region of intermediate temperatures \citep{2007ApJ...663..183P,2011ApJ...733...47W,2014A&A...567A..16S}. The significance of heating-cooling processes on the chemical balance depends on the mixing-scale of CNM/WNM structures \citep{2019ApJ...885..109B}. In the case that the characteristic CNM scale $\lambda_c$ is much less than the scale of cloud $L_c$,  the CNM and WNM are well mixed on small scales so that each cloud contains both phases. As a consequence, there would be less H$_2$ gas compared to the pure CNM case. However, when $\lambda_c\gg L_c$, the transition samples only the higher CNM densities, which is similar to our isothermal numerical setting (see \citealt{2019ApJ...885..109B} for a detailed discussion). 
	
	For the calculation of moment maps, the non-linear spectroscopic mapping between PPP space and PPV space does not require an isothermal environment (see Eq.~\ref{eq.max}). The temperature can vary from point to point in a multi-phase medium. The integration along the LOS for moment maps eliminates the effect of temperature so that the gradients are only regulated by gas density and turbulent velocity. Therefore, the moment-1 map and moment-2 map are not contaminated by thermal broadening (see also \citealt{2005ApJ...631..320E,KLP17b}). In a multi-phase scenario, the presence of gas temperature variations can complicate the velocity distributions. However, as the temperature is a passive scalar, in sub-Alfv\'{e}nic environment, we expect its pattern to be similar to the one of velocity, i.e., elongating along with the magnetic fields. In particular, the temperature is usually negatively proportional to the gas density \citep{1995ApJ...443..152W}. The temperature distribution thus can reduce the shock effect in a supersonic environment. On the other hand, the transition from WNM to CNM may occur in velocity divergence regions, which triggers the thermal instability. Consequently, the temperature differences may have a negative effect since the MHD turbulence structure may get changed. 
	\subsection{Distinguish shocks and Identify self-gravitating regions}
	
	The existence of shocks or regions of gravitational collapse is a particular case for applying GT. The GT is founded on the fact that the turbulent velocity and magnetic field eddies are elongating along the magnetic field. 
	
	As we discussed earlier, in MHD turbulence, the density statistics are different from turbulent velocity. As a result, it is not surprising that several authors have reported that the anisotropy is significantly changed by high-density filaments that form perpendicular to the magnetic field for sub-Alfv\'{e}n and highly supersonic turbulence \citep{2019ApJ...878..157X, IGs,2020MNRAS.492..668B,2020ApJ...894L...2B}.  In the density gradient's picture, its direction in front of shocks flips by 90$^\circ$, becoming parallel to the magnetic field instead of perpendicular. This change comes from the rapidly jump condition of density so that there is no particular universal density at which the change happens \citep{IGs}. At the same time, the anisotropy of the velocity field is not sensitive to shocks (see \S~\ref{sec:theory}). Therefore, shocks do not degrade the corresponding velocity gradient's performance in tracing the magnetic field, and combining velocity and density gradients, one can identify shocks.
	
	The gravitational collapse may also change the turbulence's picture. In the vicinity of gravitational collapse, the gravitational pull produces
	the most significant acceleration of the plasma in the direction parallel to the magnetic field, and the density and velocity gradients are parallel to the magnetic field \citep{Hu20}. Unlike shocks, the collapse is unbounded by the sonic Mach number $\rm M_S$, and it induces a significantly high gradient amplitude. As a result, we get the condition to distinguish shocks and identify self-gravitating regions: (i) for shocks, we have the low amplitude velocity gradient being perpendicular to the magnetic field and the low amplitude density gradient being parallel to the magnetic field; (ii) for gravitational collapse, we have both high amplitude density and velocity gradient being parallel to the magnetic field\footnote{The transfer of the parallel to perpendicular orientation happens first for density gradients and only later for velocity gradients \citep{LY18a,Hu20}. This means that density gradients change their direction at larger distances from the gravitational center. In terms of temporal evolution, the density gradients change their direction before velocity gradients.}.  The corresponding recipe for identifying the gravitational regions with the GT is presented in \citep{Hu20}.
	
	\section{Conclusion}
	\label{sec:conc}
	The photodissociation regions (PDRs) has a vital role in the formation of molecular clouds. It processes the chemical transition from atomic gas to most of the molecular gas. For example, through the essential \ion{H}{1}-to-H$_2$ transition, the formation of other heavy molecules is achievable. The role of turbulence and magnetic field in PDRs is not well understood, and it requires further studies. We explore the possibility of studying magnetic fields of the PDRs by performing MHD simulation with numerical cubes in \ion{H}{1}-H$_2$ chemical balance. We study the properties of turbulence by calculating the energy spectrum and structure-function. This work also suggests the promise of using multiple moment maps to probe magnetic fields. We summarize our results as follows:
	\begin{enumerate}
		\item Using the synthetic \ion{H}{1} and H$_2$ cubes in chemical balance, we confirm that:
		\begin{enumerate}
			\item The density fluctuation induced by turbulence can disperse the distribution of H$_2$ and \ion{H}{1} fraction. 
			\item The energy spectrum of moment maps gets shallower with an increment of $\rm M_s$.
		\end{enumerate}
		\item Analogy with IGs and VCGs, we introduce the gradients of moment-2 map, denoted as Velocity Square Gradients (VC$_2$Gs), to trace the magnetic fields. In particular, we show that in an isothermal environment:
		\begin{enumerate}
			\item VC$_2$Gs are more accurate than VCGs and IGs in tracing the magnetic field orientation. 
			\item VC$_2$Gs, VCGs, and IGs tend to be parallel to the local magnetic fields when getting close to the dense shock front in the absence of gravity. 
			\item VC$_2$Gs are less sensitive to shocks than IGs. VC$_2$Gs have the advantage of tracing the magnetic field in supersonic turbulence.
		\end{enumerate}
		\item We show that the synergy of the Velocity Gradients Technique (VGT) and Principal Component Analysis improves the accuracy of magnetic field tracing.
	\end{enumerate}
	
	\acknowledgments
	Y.H. acknowledges the support of the NASA TCAN 144AAG1967. A.L. acknowledges the support of the NSF grant AST 1715754, and 1816234. 
	S.B acknowledges support from the Harvard-Smithsonian Institute for Theory and Computation (ITC).
	The Flatiron Institute is supported by the Simons Foundation. 
	
	%
	
	\vspace{10mm}
	
	\software{Julia \citep{2012arXiv1209.5145B}, ZEUS-MP/3D \citep{2006ApJS..165..188H}, Paraview \citep{ayachit2015paraview}}

	


	
	
\end{document}